\documentclass[useAMS]{mn2e}
\usepackage{epsfig,lscape}
\usepackage{graphicx}

\DeclareGraphicsExtensions{.eps,.eps.gz,.epsi}
\newcommand\apj{ApJ}%
\newcommand\apjl{ApJL}%
\newcommand\aap{A\&A}%
\newcommand\pasp{PASP}%
\newcommand\mnras{MNRAS}%
\newcommand\jcap{JCAP}%
\newcommand\apjs{ApJS}%
\newcommand\aaps{A\&AS}%

\newcommand{\eld}{${\it N}_{\rm e}$}

\newcommand{\elt}{${\it T}_{\rm e}$}

\newcommand{\teff}{${\it T}_{\rm eff}$}

\newcommand{\cmt}{cm$^{-3}$}
\newcommand{\tsq}{$t^{2}$}

\newcommand{\foiii}{[O~{\sc iii}]}
\newcommand{\fniii}{[N~{\sc iii}]}
\newcommand{\foi}{[O~{\sc i}]}
\newcommand{\foii}{[O~{\sc ii}]}
\newcommand{\foiv}{[O~{\sc iv}]}
\newcommand{\fsii}{[S~{\sc ii}]}
\newcommand{\fsiii}{[S~{\sc iii}]}
\newcommand{\fsiv}{[S~{\sc iv}]}
\newcommand{\fci}{[C~{\sc i}]}
\newcommand{\fni}{[N~{\sc i}]}
\newcommand{\fnii}{[N~{\sc ii}]}

\newcommand{\fariv}{[Ar~{\sc iv}]}
\newcommand{\fcliii}{[Cl~{\sc iii}]}

\newcommand{\fneii}{[Ne~{\sc ii}]}
\newcommand{\fneiii}{[Ne~{\sc iii}]}

\newcommand{\fnev}{[Ne~{\sc v}]}

\newcommand{\nii}{N~{\sc ii}}
\newcommand{\niii}{N~{\sc iii}}

\newcommand{\oii}{O~{\sc ii}}
\newcommand{\cii}{C~{\sc ii}}
\newcommand{\neii}{Ne~{\sc ii}}
\newcommand{\mgii}{Mg~{\sc ii}}
\newcommand{\ciii}{C~{\sc iii}}
\newcommand{\civ}{C~{\sc iv}}

\newcommand{\fariii}{[Ar~{\sc iii}]}

\newcommand{\hi}{H\,{\sc i}}

\newcommand{\hei}{He~{\sc i}}
\newcommand{\heii}{He~{\sc ii}}

\newcommand{\ha}{H$\alpha$}
\newcommand{\hb}{H$\beta$}
\newcommand{\hg}{H$\gamma$}

\title[3D photoionization models of NGC\,6153]{ Three-dimensional chemically homogeneous and bi-abundance photoionization 
models of the ``super-metal-rich'' planetary nebula NGC\,6153 }
\author[H.-B. Yuan et al.]{H.-B. Yuan$^1$,  
X.-W. Liu$^{1,2}$\thanks{E-mail: x.liu@pku.edu.cn}, D. P\'{e}quignot$^3$, R. H. Rubin$^{2,4,5}$, B. Ercolano$^{6,7,8}$ 
\newauthor and Y. Zhang$^9$\\
\\
$1$ Department of Astronomy, Peking University, Beijing 100871, P. R. China\\
$2$ Kavli Institute for Astronomy and Astrophysics, Peking University, 
Beijing 100871, P. R. China\\
$3$ LUTH, Laboratoire l'Univers et ses Th\'eories, associ\'e au CNRS (FRE 2462) et
\'a l'Universit\'e Paris 7, Observatoire de Paris-Meudon,\\
F-92195 Meudon C\'edex, France\\
$4$ NASA/Ames Research Center, Moffett Field, CA 94035-1000, USA\\
$5$ Orion Enterprises, M.S. 245-6, Moffett Field, CA 94035-1000, USA\\
$6$ School of Physics, University of Exeter, Stocker Road, Exeter, EX4 4QL, UK\\
$7$ Department of Physics and Astronomy, University College London, Gower Street, London, WC1E 6BT, UK\\
$8$ Institute of Astronomy, University of Cambridge, Madingley Road, Cambridge, CB3 OHA, UK\\
$9$ Department of Physics, University of Hong Kong, Pokfulam Road, Hong Kong}
\begin{document}
\date{Received:}

\pagerange{\pageref{firstpage}--\pageref{lastpage}} \pubyear{2010}

\maketitle

\label{firstpage}

\begin{abstract}{
Deep spectroscopy of the planetary nebula (PN) NGC\,6153 shows that its heavy element 
abundances derived from optical recombination lines (ORLs) are 
ten times higher than those derived from collisionally excited lines (CELs),
and points to the existence of H-deficient inclusions embedded in the diffuse nebula. 
In this study, we have constructed chemically homogeneous and bi-abundance three-dimensional photoionization models,
using the Monte Carlo photoionization code {\sc mocassin}.
We attempt to reproduce the multi-waveband spectroscopic and imaging observations
of NGC\,6153, and investigate the nature and origin of the postulated H-deficient inclusions, as well as 
their impacts on the empirical nebular analyses assuming a uniform chemical composition.
Our results show that chemically homogeneous models yield small
electron temperature fluctuations and fail to reproduce the
strengths of ORLs from C, N, O and Ne ions. In contrast, bi-abundance models
incorporating a small amount of metal-rich inclusions ($\sim 1.3$
per cent of the total nebular mass) are able to
match all the observations within the measurement uncertainties.
The metal-rich clumps, cooled down to a very low temperature ($\sim 800$~K) by 
ionic infrared fine-structure lines, dominate the emission of 
heavy element ORLs, but contribute almost nil to the emission of most CELs.
We find that the abundances of C, N, O and Ne derived empirically from CELs, assuming a uniform chemical composition,
are about 30 per cent lower than the corresponding average values of the whole nebula, including the contribution
from the H-deficient inclusions.
Ironically, in the presence of H-deficient inclusions, 
the traditional standard analysis of the optical  helium recombination lines,
assuming a chemically homogeneous nebula, overestimates the helium abundance by 40 per cent.
}
\end{abstract}


\section{Introduction}
\label{intro}

A long-standing dichotomy in nebular astrophysics is that
the heavy element abundances derived from optical recombination lines (ORLs)
are systematically higher than those derived from collisionally excited 
lines (CELs; see Peimbert \& Peimbert 2006 and Liu 2006 for recent reviews).
Temperature fluctuations (Peimbert 1967), density inhomogeneities (Rubin 1989; 
Viegas \& Clegg 1994) and X-ray irradiation of quasineutral chemically homogeneous clumps (Ercolano 2009) 
have been proposed to explain this discrepancy. However,
extensive observations  of planetary nebulae (PNe) have shown that they are 
incapable of accounting for all observational features (Liu 2006). 
In order to explain the CEL/ORL abundance dichotomy, Liu et al. (2000, L2000 hereafter)
presented a bi-abundance model. The model assumed that the nebula contains two gas components of different 
abundances --- the diffuse component of ``normal'' composition that  dominates the emission of CELs and
another consisting of cold H-deficient inclusions posited in the diffuse gas that dominates the emission of ORLs. 
The model explains well spectroscopic observations from the ultraviolet (UV) to the infrared (IR). 
However, detailed photoionization models are still required to 
test this scenario and address issues such as: `How can the nebular physical conditions and its elemental 
abundances be reliably determined in the exiseence of chemical inhomogeneities?', 
and `What is the nature and origin of the H-deficient
clumps?' To address these issues, this paper presents the results of
three-dimensional photoionization modelling for the  PN NGC\,6153.

NGC\,6153 is an archetypal nebula showing a particularly large ORL versus CEL abundance discrepancy 
factor (adf) of about ten (L2000). A large number of ORLs from a variety of heavy-element ions have been
detected in this PN. NGC\,6153 is thus an ideal object to model and to investigate the possible causes of the  
CEL/ORL abundance discrepancy problem. Early optical studies by Pottasch et al. (1984, 1986) suggested that NGC\,6153
has a very high metallicity. The result was further corroborated by the observations of the Short Wavelength 
Spectrometer on board the Infrared Space Observatory ({\it ISO}-SWS) in 
the mid-IR ( Pottasch et al. 2003). L2000 presented a comprehensive
spectroscopic analysis of NGC\,6153 from the UV to the far-IR and found that the C, N, O, and Ne 
abundances derived from ORLs are about ten times higher than those derived from CELs. 
Using simple empirical models, L2000 showed that the discrepancy can be explained by assuming 
the existence of a H-deficient component embedded in the diffuse gas.
Using optical integral field spectroscopic observations of three PNe including NGC\,6153  with the Very Large Telescope 
Fibre Large Array Multi Element Spectrograph Argus integral field unit, Tsamis et al. (2008) provided further evidence 
for the existence of cold H-deficient inclusions and constrained their physical sizes to be smaller than $\sim$1,000 astronomical units. 
P\'{e}quignot et al. (2002, 2003) constructed one-dimensional photoionization models
that contain two gas components of different abundances. The models reproduced satisfactorily
most of the observed spectral features of NGC\,6153. 
It is however difficult to self-consistently treat the diffuse radiation fields 
in one-dimensional models and investigate the size and spatial distribution of the 
postulated H-deficient clumps.
This motivated us to carry out further modelling using the 
Monte Carlo photoionization code {\sc mocassin} (Ercolano et al. 2003a, 2005, 2008) capable of dealing
with an arbitrary nebular geometry and density and abundance distributions.

The paper is organized as follows. In Section 2, previous modelling work and
observational data used to constrain our models are summarized.
In Section 3, we introduce the code used for the modelling and present the modelling results. 
In Section 4, we discuss the implications and limitation of our results, the properties of the 
H-deficient inclusions and their impacts on nebular studies.  
A summary is given in Section 5.

\section{Observations and previous models}
\label{obs}
In this section, we summarize observations available to constrain the
photoionization models of NGC\,6153  and introduce the earlier one-dimensional bi-abundance work
of P\'{e}quignot et al. (2002, 2003), which serves as the starting point of our new three-dimensional models.

\subsection{Spectroscopic observations}

Spectra of NGC\,6153 from the UV to the far-IR are available from the literature (e.g.,
L2000, Pottasch et al. 1984, 1986, 2003).  For a given wavelength range,
there may exist more than one set of observations. Due to 
varying aperture sizes, orientations and measurement uncertainties, different studies
often obtain different fluxes for a given emission line. For the current work, we have adopted 
line fluxes of L2000, who present measurements in the optical obtained with ground-based facilities, in the 
UV obtained with the International Ultraviolet Explorer ({\it IUE}; 1150--3300{\AA}), and 
in the mid- to far IR obtained with the Infrared Astronomical Satellite
({\it IRAS};7.6--22.7\,$\mu$m) and the Long Wavelength Spectrometer on board the {\it ISO} ({\it ISO}-LWS; 43--197\,$\mu$m). 
L2000 obtained two sets of optical spectra. The set of spectra with the slit oriented
along the nebular minor axis allows one to investigate the spatial brightness
distribution of a given line along the slit. The other set, obtained by 
scanning the slit across the whole nebular surface,  when combined with the
total H$\beta$ flux [$\log$\,$F$(\hb)~= $-$10.86\,ergs~cm$^{-2}$s$^{-1}$; Cahn et al. 1992],
yields integrated fluxes of the whole nebula for all the detected lines.
The integrated line fluxes can be directly compared with the model predictions.
Note that L2000 used a logarithmic extinction, {\it c}(\hb)~=~1.30,
deduced from the Balmer decrement, {\bf and the Galactic reddening law of Howarth (1983)} to deredden the UV, optical and near-IR spectra.
They adopted a different value {\it c}(\hb)~=~1.19, deduced from the observed radio continuum flux
density and the \hb~flux, to deredden the mid- and far-IR spectra.
The same dereddening algorithm has been used in the current study.
Here we have added a few mid-IR CELs falling within the 
spectral window of the {\it ISO}-SWS 
in our modelling. Given the fact that the size of the {\it ISO}-SWS diaphragm was 
not large enough to cover the entire nebula and the telescope pointing was offset by $3\farcs49$ east 
and $9\farcs0$ south of the nebular centre, aperture corrections have been applied by comparing the line fluxes 
obtained with {\it ISO}-SWS and those with {\it IRAS}.
All strong emission lines considered in the previous modelling (L2000, P\'{e}quignot et al. 2003) were included in the current work, 
c.f. the next section for details.

In addition to line fluxes from the literature, we have supplemented the data with new observations using the 
B\&C spectrograph mounted on the European Southern Observatory (ESO) 1.52 m telescope. 
The spectra, obtained on 2001 May 12,  were taken in order to measure several near-IR emission 
lines (He~{\sc i} $\lambda$10830, {\fci} $\lambda$9850, and {\fsiii} $\lambda$$\lambda$9069,9531), 
important for model constraints.  The He~{\sc i} $\lambda$10830 line is particularly important to constrain the 
helium abundance in a bi-abundance model (c.f. Section 4.1). The spectra covered the
wavelength range 7700--13254\,{\AA}, with the slit placed along the nebular minor axis.
Two long exposures, each of 1800\,s, and one short exposure of 300\,s were made.
The spectra were reduced using the {\sc long92} package in {\sc midas}
\footnote{ {\sc midas} is developed and distributed by the European Southern Observatory.} following the standard procedures.
The spectra were bias-subtracted, flat-fielded, cosmic rays cleaned and
wavelength-calibrated using exposures of a He-Ar calibration lamp. They were then flux-calibrated
with the {\sc iraf}\footnote{ {\sc iraf} is distributed by the National Optical Astronomy Observatory, 
which is operated by the Association of Universities for Research in Astronomy (AURA) under cooperative agreement with 
the National Science Foundation.} package using wide-slit observations of {\it HST} standard stars.
After dereddening, the spectra were normalized such that
\hi~{\it I}(P12)/{\it I}(\hb)~=~0.01106, as predicted by the recombination theory, Case\,B (Storey \& Hummer 1995)
for \elt~=~6,080 {\rm K} and \eld~=~3,500 {\rm cm}$^{-3}$. The \elt~and \eld~values here are taken from L2000. 
Fluxes deduced from these observations are tabulated in Table~1 (see Section~3).

In an attempt to detect directly the postulated H-deficient knots and to constrain their physical
sizes, we have also carried out long-slit spectroscopy of NGC\,6153 using the STIS instrument on board the Hubble Space Telescope ({\it HST}).
On 2002 June 3--4, two gratings, G430M and G750M, were used to cover the spectral range 
$\lambda$$\lambda$3050-5600 and $\lambda$$\lambda$5450-10100, respectively.
A long slit of $52\arcsec\times0\farcs2$  was placed along the nebular minor axis through the central star to
cover the brightest parts of the nebula. The spatial resolution along the slit was $0\farcs05$/pixel.
The standard pipeline procedures in {\sc iraf/stsdas} were used to reduce the data. 
Unfortunately, these data suffered from  poor signal-to-noise ratio  as well as severe contamination
by cosmic rays and turned out to be of very limited value for the current investigation (see Section 4.4).

\begin{table*}
\label{modelresults}
\centering
\caption{Comparison of the model predictions and the observations. The
observed intensities have been dereddened and normalized such that $I$(\hb)$=100$.
Columns (4)--(12) give the ratios of predicted over observed values (departure ratios).  }
\begin{tabular}{lrrccccccccc}
\noalign{\vskip3pt} \noalign{\hrule} \noalign{\vskip3pt}
Line    &$\lambda$ (\AA)$^a$ &$I_{\rm obs}$   &S     &E1    &E2    & B$_{\rm n}$    &B$_{\rm c}$    & B & B$'_{\rm n}$    &B$'_{\rm c}$    & B$'$\\
\noalign{\vskip3pt} \noalign{\hrule}\noalign{\vskip3pt}     
                                                            
\multicolumn{9}{c}{H, He recombination lines, optical and radio continua}\\

\hb     & 4861.3        &100        &1.01     &1.01     &1.00      &0.95 &0.06 &1.01 &0.94&0.06&1.00\\
\ha     & 6562.8        &298.08     &0.97     &0.97     &0.97       &0.90 &0.08 &0.98 &0.90 &0.08 &0.98\\
\hg     & 4340.5        & 48.92     &0.95     &0.95     &0.95      &0.89 &0.06 &0.95 &0.89 &0.06 &0.95\\
\noalign{\vskip2pt}
\hei    & 3888.6        & 11.88     &1.10     &1.07     &1.06      &0.78 &0.17 &0.95 &0.78 &0.17 &0.95 \\
\hei    & 4471.5        & 6.47      &1.02     &1.02     &1.02      &0.68 &0.27 &0.95 &0.68 &0.27 &0.95\\
\hei    & 5875.7        &18.85      &1.01     &1.02     &1.02      &0.67 &0.35 &1.02 &0.68 &0.34 &1.02\\ 
\hei    & 6678.2        & 4.83      &1.08     &1.09     &1.09      &0.72 &0.40 &1.12 &0.72 &0.39 &1.11\\ 
\hei    & 7065.2        & 4.35      &1.23     &1.37     &1.34      &0.86 &0.11 &0.97 &0.86 &0.11 &0.97\\ 
\hei    & 7281.6        & 0.56      &1.89     &1.91     &1.91      &1.32 &0.23 &1.55 &1.32 &0.23 &1.55\\ 
\hei    &10830.3        & 59.96     &1.59     &1.79     &1.74      &1.21 &0.10 &1.31 &1.21 &0.10 &1.31\\
\heii   & 1640.0        & 82.19     &1.03     &0.98     &0.99      &0.86 &0.13 &0.99 &0.81 &0.21 &1.02\\
\heii   & 4686.0        & 12.85     &1.02     &0.97     &0.98      &0.84 &0.13 &0.97 &0.80 &0.21 &1.01\\
\noalign{\vskip2pt}
BJ/\hb  & 3646          &0.60       &0.8     &0.8     &0.8      &0.75     &0.26     &1.01 &0.75&0.26&1.01\\
\noalign{\vskip2pt}
\multicolumn{9}{c}{Heavy-element recombination lines}\\
\cii    & 4267.2        &2.42       &0.16     &0.16     &0.16      &0.10 &0.85 &0.95 &0.10 &0.82 &0.92\\ 
\ciii   &4650+(M1)          &0.50       &1.11     &0.76     &0.76      &0.41 &0.27 &0.68 &0.40 &0.39 &0.79\\
\ciii   &4187(M18)          &0.08       &1.45     &1.00     &1.00      &0.54 &1.1  &1.64 &0.52 &1.60 &2.12\\
\noalign{\vskip2pt}
\nii    &4041.3        &0.24       &0.19     &0.21     &0.21      &0.15 &0.89 &1.04 &0.16 &0.84 &1.00\\
\nii    &4241.8        &0.22       &0.14     &0.15     &0.15      &0.11 &0.66 &0.77 &0.11 &0.62 &0.73\\
\nii    &5679.0+(V3)        &0.94       &0.26     &0.29     &0.28      &0.21 &0.72 &0.93 &0.21 &0.69 &0.90\\ 
\niii   &4379.0+(M17)        &0.66       &0.60     &0.45     &0.46      &0.30 &0.94 &1.24 &0.30 &1.35 &1.65\\
\noalign{\vskip2pt}
\oii    &4075.0+(V10)        &4.12       &0.15     &0.15     &0.16      &0.10 &0.64 &0.74 &0.10 &0.67 &0.77\\
\oii    &4089.3        &0.55       &0.13     &0.13     &0.14      &0.09 &0.83 &0.92 &0.09 &0.86 &0.95\\ 
\oii    &4649.1        &1.39       &0.23     &0.23     &0.24      &0.16 &0.89 &1.05 &0.16 &0.92 &1.08\\
\oii    &4651.0+(V1)        &3.68       &0.21     &0.21     &0.23      &0.15 &0.84 &0.99 &0.15 &0.87 &1.02\\ 
\noalign{\vskip2pt}
\neii   &3340.0+(V2)        &1.96       &0.09     &0.09     &0.09      &0.08 &0.56 &0.64 &0.08 &0.61 &0.69\\
\neii   &3700.0+(V1)       &1.07       &0.10     &0.09     &0.10      &0.08 &0.54 &0.62 &0.08 &0.60 &0.68\\
\neii   &4392.0         &0.15       &0.11     &0.11     &0.11      &0.06 &0.56 &0.65 &0.09 &0.61 &0.70\\
\noalign{}
\mgii   &4481.0         &0.05       &         &         &          &0.66 &0.26 &0.92 &0.67 &0.27 &0.93\\ 
\noalign{\vskip2pt}
\multicolumn{9}{c}{Collisionally excited lines (UV and optical)}\\
\fci        &9850.3     &0.17       &0.06     &0.17     &0.34      &0.25 &0.07 &0.32 &0.26 &0.05 &0.31\\
C~{\sc iii}]&1907.0+      &46.48      &1.22     &1.28     & 1.06     &0.90 &0.07 &0.97 &0.87 &0.11 &0.98\\ 
\civ       &1549.0+       & $<$30.0   &1.08     &0.61     & 0.61     &0.53 &0.00 &0.53 &0.49 &0.00 &0.49\\
\noalign{\vskip2pt}
\fni       &5199.0+     & 0.24      &0.02     &0.05     &1.06      &1.16 &0.06 &1.22 &1.12 &0.04 &1.16\\ 
\fnii      &5754.6      & 0.83      &0.30     &0.61     &0.75      &1.03 &0.21 &1.24 &1.08 &0.20 &1.28\\ 
\fnii      &6583.5+     & 64.20     &0.31     &0.72     &0.99      &1.08 &0.05 &1.13 &1.11 &0.05 &1.16\\ 
N~{\sc iii}] &1744.0+     &16.30      &0.32     &0.44     &0.33      &0.42 &0.00 &0.42 &0.40 &0.00 &0.40\\ 
\noalign{\vskip2pt}
\foi       &6300.3      &0.78       &0.00     &0.02     &0.74      &0.64 &0.00 &0.64 &0.69 &0.00 &0.69\\ 
\foii      &3726.1      &18.9       &0.40     &0.75     &0.89      &0.86 &0.22 &1.08 &0.90 &0.25 &1.15\\ 
\foii      &3728.8      &9.69       &0.47     &0.80     &0.91      &0.94 &0.20 &1.14 &0.97 &0.22 &1.18\\ 
\foiii     & 4363.2     & 4.19      &0.90     &1.05     &0.93      &0.92 &0.00 &0.92 &0.88 &0.01 &0.89\\ 
\foiii     & 5006.8+    &1189.90    &1.10     &1.19     &1.16      &0.96 &0.00 &0.96 &0.94 &0.00 &0.94\\ 
\noalign{\vskip2pt}
\fneiii    &3869.0     &93.91      &0.96     &1.00     &0.95      &1.01 &0.00 &1.01 &0.99 &0.00 &0.99\\ 
\noalign{\vskip2pt}
 \fsii     &4068.6     &1.03       &0.13     &0.31     &0.52      &0.56 &0.00 &0.56 &0.59 &0.00 &0.59\\
\fsii      &4076.4     &0.35       &0.13     &0.31     &0.52      &0.55 &0.00 &0.55 &0.58 &0.00 &0.58\\
\fsii      &6716.4     &3.16       & 0.22    &0.38     &0.57      &0.65 &0.00 &0.65 &0.65 &0.00 &0.65\\
\fsii      &6730.8     &5.26       & 0.19    &0.36     &0.56      &0.61 &0.00 &0.61 &0.62 &0.62 &0.62\\ 
\fsiii     &6312.1     &1.29       &0.97     &1.33     &1.16      &1.39 &0.00 &1.39 &1.38 &0.00 &1.38\\
\fsiii     &9068.9     &39.74      &0.86     &1.10     &1.03      &1.09 &0.00 &1.09 &1.09 &1.09 &1.09\\
\fsiii     &9531.0     &113.68     &0.74     &0.95     &0.89      &0.95 &0.00 &0.95 &0.95 &0.00 &0.95\\
\noalign{\vskip2pt}
\fcliii    &5517.7      &0.55       & 1.07    &1.13     &1.01      &1.08 &0.00 &1.08 &1.07 &0.00 &1.07\\
\fcliii    &5537.7      &0.70       & 0.88    &1.01     &0.90      &0.95 &0.00 &0.95 &0.94 &0.00 &0.94\\
\noalign{\vskip2pt}
\fariii    &5191.8      &0.10       &0.83     &0.84     &0.77      &0.94 &0.00 &0.94 &0.91 &0.00 &0.91\\
 \fariii   &7135.8     &19.00       &1.00     &0.96     &0.97      &1.01 &0.00 &1.01 &1.00 &0.00 &1.00\\
\fariv     &4711.4      &2.54       &1.00     &0.97     &0.98      &0.95 &0.00 &0.95 &0.93 &0.00 &0.93\\
\fariv     &4740.2      &2.35       &1.00     &0.99     &0.99      &0.96 &0.00 &0.96 &0.93 &0.00 &0.93\\
\noalign{\vskip2pt}
\multicolumn{9}{c}{Collisionally excited lines (IR)}\\
\fniii   &57.3\,$\mu$m  &72.43      &1.14     &1.03     &1.02      &0.85 &0.16 &1.01 &0.86 &0.16 &1.02\\
\noalign{\vskip2pt}
\foiii  &51.8\,$\mu$m   &267.60     &0.91     &0.79     &0.85      &0.62 &0.23 &0.85 &0.63 &0.25 &0.88\\
\foiii  &88.4\,$\mu$m   &74.44      &0.84     &0.60     &0.65      &0.49 &0.13 &0.62 &0.49 &0.14 &0.63\\
\end{tabular}
\end{table*}

\setcounter{table}{0}
\begin{table*}
\centering
\caption{\it --continued} 
\begin{tabular}{lrrccccccccc}

\noalign{\vskip3pt} \noalign{\hrule} \noalign{\vskip3pt}
Line    &$\lambda$ (\AA)$^a$ &$I_{\rm obs}$   &S     &E1    &E2    & B$_{\rm n}$    &B$_{\rm c}$    & B & B$'_{\rm n}$    &B$'_{\rm c}$    & B$'$\\
\noalign{\vskip3pt} \noalign{\hrule} \noalign{\vskip3pt}
\foiv   &25.9\,$\mu$m   &118.10     &1.18     &1.18     &1.29      &1.10 &0.16 &1.26 &1.04 &0.29 &1.33\\
\fneii  &12.8\,$\mu$m   &23.14      & 0.07    &0.11     &0.15      &0.16 &1.00 &1.16 &0.15 &0.68 &0.83\\
\fneiii &15.6\,$\mu$m   &253.50     &1.00     &0.94     &1.00      &0.88 &0.50 &1.38 &0.88 &0.59 &1.47\\
\fneiii &36.0\,$\mu$m   & 32.61     & 0.65    &0.60     &0.64      &0.57 &0.18 &0.75 &0.57 &0.21 &0.78\\
\noalign{\vskip2pt}
\fsiii &18.7\,$\mu$m    &47.28      &1.03     &1.20     &1.18      &1.18 &0.03 &1.21 &1.18 &0.03 &1.21\\
\fsiii &33.5\,$\mu$m    &32.61      &0.81     &0.78     &0.77      &0.80 &0.02 &0.82 &0.80 &0.01 &0.81\\
\fsiv  &10.5\,$\mu$m    &362.16     &1.09     &0.94     &0.93      &0.87 &0.01 &0.88 &0.87 &0.01 &0.88\\
\noalign{\vskip2pt}
\fariii & 9.0\,$\mu$m  &56.34      &0.40     &0.37     &0.39      &0.37 &0.01 &0.38 &0.37 &0.01 &0.38\\
\fariii &21.8\,$\mu$m   &2.69       &0.53     &0.48     &0.51      &0.49 &0.00 &0.49 &0.49 &0.00 &0.49\\
\noalign{\vskip3pt} \noalign{\hrule}\noalign{\vskip3pt}
\end{tabular}
\begin{description}
\item[$^a$] `+' refers to the total intensity of the multiplet.
\end{description}
\end{table*}

\subsection{Imaging observations}
\label{image}
Imaging observations of NGC\,6153 have been carried out with ground-based telescopes (Pottasch et al. 1986) 
and the {\it HST} (L2000). 
Fig.~1 shows the {\it HST} images taken under programme GO-8594 (PI Liu) with the 
WFPC2 camera in the F502N and F656N narrow-band filters on
2000 August 13.\footnote{Based on observations made with the
NASA/ESA {\it Hubble Space Telescope}, obtained at the Space Telescope
Science Institute, which is operated by AURA, Inc., under NASA contract NAS5-26555.}
Two 600s and two 500s exposures were made in the F502N and F656N filters, respectively.
The pixel size was $0\farcs0455$. The images were co-added, cosmic-rays 
removed and calibrated following the standard recipes (e.g. Holtzmann et al. 1995). They were then  dereddened
using an extinction coefficient $c({\rm H}\beta)=1.30$.  
The two filters trace respectively emission of the \foiii~$\lambda$5007 and H$\alpha$ lines, although
there could be some contamination from the nebular continuum emission and the  \fnii~$\lambda$$\lambda$6584,6548 lines 
in the case of F656N. We performed  aperture photometry and found that
\ha~and \foiii~$\lambda$5007 line fluxes deduced from {\it HST} images
agreed within 10 per cent with the  corresponding values yielded by the optical scanning spectrum of L2000, 
suggesting insignificant contamination to the  {\it HST} F656N image from the \fnii~lines and the nebular continuum emission.

The azimuthally averaged radial surface brightness distribution of \ha~was deduced from
the {\it HST} F656N image and plotted in Fig.~2. The profile is used in our models to
constrain the nebular density distribution.

\begin{figure*}
\label{sphfluxdist}  
 \centering
 \epsfig{file=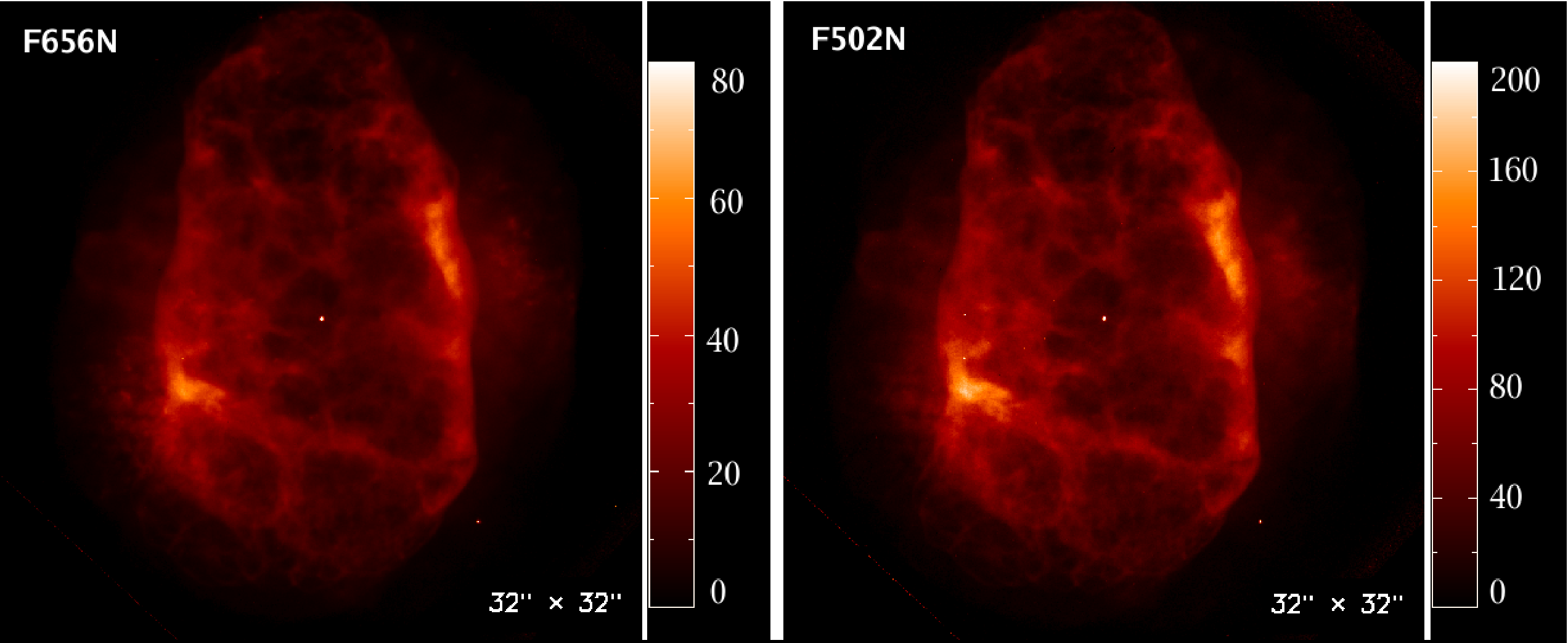, width=16cm} 

 \caption{{\it HST}/WFPC2 images of NGC\,6153 obtained with the F656N (left panel) and F502N (right panel) narrow-band filters. 
North is up and east to the left.
The units of the color bars are $10^{-13} ~{\rm ergs~ cm}^{-2}~ {\rm s}^{-1} ~{\rm arcsec}^{-2}$.}
\end{figure*}

\subsection{Previous models}

Previous empirical and one-dimensional photoionization models of NGC\,6153 
provide valuable insight and a useful guide for setting the initial parameters of the new
3D photoionization models.
L2000 constructed several parametrized empirical models to match the
observed fluxes of selected ORLs and CELs. Recombination contributions to CELs 
were considered. They obtained the following results: 1)
Models of a uniform chemical composition and density distribution cannot reproduce the
large strengths of ORLs; 2) Chemically homogeneous models with density variations
predict too weak ORLs and the hydrogen Balmer discontinuity and a too strong [Ne~{\sc iii}] fine-structure line
compared to observations; and 3) Bi-abundance models consisting of two components, each 
with a different temperature, density, and chemical composition, can account for most of the 
observed patterns of NGC\,6153. Two different bi-abundance models were presented by L2000.
One assumed some low-density (\eld $~$= 700 cm$^{-3}$) and low-temperature (\elt $~$= 500 K)
H-deficient material embedded in the ``normal'' nebular gas of
\eld$~$= 5,500 cm$^{-3}$ and  \elt $~$= 9,500 K. In the other model,
the hydrogen-depleted material was assumed to be fully ionized and have a high density 
($2\times 10^6$ cm$^{-3}$) and a moderate temperature of 4,700\,K. 
In both models, the helium and heavy element abundances with respect to hydrogen in
the H-deficient component were respectively $~4$ and$~100$ times higher than the corresponding values 
in the diffuse gas.

P\'{e}quignot et al. (2002, 2003) constructed detailed bi-abundance photoionization models of NGC\,6153 using
the one-dimensional code {\rm NEBU}. Their models were composed of sectors extracted from 
a number of spherically symmetric models, each  made of different input parameters
except the central ionizing source. In the models, a H-deficient component
with a small filling factor was mixed with a normal-composition shell. 
They found that models containing a small amount of dense, cold H-deficient material 
[$M$ = 0.0031$\mathrm{{\it M}_\odot}$, ${\it N}({\rm H})$ = 4\,410 \cmt, \elt$~$= 1\,390 K] in pressure equilibrium with the 
 surrounding gas of  ``normal'' composition [$M$ = 0.38$\mathrm{{\it M}_\odot}$, ${\it N}{\rm (H)}$ = 1\,170 \cmt, \elt$~$= 9\,040 K]  
can well account for the observed fluxes of most  ORLs and CELs. Compared to
the normal gas, the H-deficient component is enhanced in helium and CNONe elements by factors of 6
and 100, respectively. Nevertheless, the average elemental abundances of the entire nebula are only slightly affected
by the presence of the H-deficient component, given its small mass.
P\'{e}quignot et al. (2002) pointed out that helium abundance relative to hydrogen will be overestimated 
if composition fluctuations are not taken into account and it is also essential to consider collisional contribution 
to excitation of \hei~lines to correctly determine helium abundance.
\section{Models}
\label{models}

The modelling is carried out using {\sc mocassin} (Version 2.02.12; Ercolano et al. 2003a, 2005), a fully 
three-dimensional Monte Carlo photoionization code that solves the radiative transfer
self-consistently in an iterative way and is designed to tackle problems involving density variations and
chemical inhomogeneities. The code is therefore suitable for the current investigation aiming to study the 
effects of the complicated geometry and possible chemical inhomogeneities of NGC\,6153.
To reduce degree of freedom, the ionizing star is assumed to have a blackbody spectral energy distribution in our models.
A more realistic representation of the ionizing star is left to a later work.
The nebula is approximated by a cubical Cartesian grid of cells of individually pre-defined density and chemical composition. 
For each cell, the radiation fields (including the stellar and the diffuse component) 
and physical conditions (electron temperature, density and ionic fractions of all elements considered) 
are calculated iteratively. 
Based on an initial guess of the temperature and ionization structures, the code calculates radiation 
fields, solves the ionization and thermal equilibrium equations and then updates the 
temperature and ionization structures. This process is iterated until physical conditions are converged in most cells.
Finally, the emission spectrum of the model nebula is calculated by integrating over the volume of the nebula.
The code has been applied to construct photoionization models of the PN NGC\,3918,     
the H-deficient knots in the ``born-again'' PN Abell\,30, the Wolf-Rayet PN NGC\,1501 (Ercolano et al. 2003b,c, 2004)
and PN NGC\,6781 (Schwarz \& Monteiro, 2006).

In our models, we have assumed that NGC\,6153 has an azimuthally symmetrical structure in the x and y plane 
and a reflection symmetry in Z and placed the central star at the centre of a corner cell of a $48\times48\times48$
cubic grid.  Modelling in this way can effectively save the computational time 
since only one eighth of the nebula is simulated.
The  {\it HST} images of NGC\,6153 show that the nebula can be approximated by 
an ellipsoidal shell with enhanced emission in the  equator.  The brightest part of the nebula
exhibits point symmetry with respect to the central star, probably implying the existence of 
an ellipsoidal ring lying along the major axis in the south-east to the north-west direction.
A faint halo can also be seen surrounding the bright nebula. 
In order to better constrain the input parameters,
we have constructed a series of models with progressively increasing complexity. These  include
a spherical model (model S), two ellipsoidal models without and with a torus (models E1 and E2, respectively), 
and a bi-abundance model (model B). The spherical and ellipsoidal models are chemically homogeneous. 
Table~2 gives the adopted model parameters that yield the best match to the observations. In Table~1,
we compare the predicted line fluxes with the observations. Further details of our models are presented below.

\setcounter{table}{1}
\begin{table}
  \centering
  \caption{Model parameters.}\label{modelparameters}

  \begin{tabular}{|l|l|l|l|l|l|l|}
    \hline
        Parameter                          &S     &E1     &E2     &B$_{\rm n}$     &B$_{\rm c}$     &B    \\
\hline
{\it L} (e36 ergs s$^{-1}$)                 &16    &14     &14     &13     &13     &13   \\
 \teff\ (kK)                               &90    &92     &92     &92     &92     &92   \\   
{\it M} (M$_{\odot}$)                      &0.349 &0.269  &0.267  &0.243  &0.0031 &0.246\\ 
${\it N}_{\rm tot}$(H) (e56)               & 2.63 &2.03   &2.02   &2.07   &0.0081 &2.08\\
       He                                  &0.138 &0.138  &0.138  &0.10   &0.50   &0.102\\
       C (e$-$4)                           &5.4   &4.9    &4.9    &3.2    &177    &3.88\\
       N (e$-$4)                           &4.6   &4.6    &4.6    &3.8   &150    &4.37\\
       O (e$-$4)                           &6.82  &6.82   &7.44   &5.63   &440    &7.33\\
       Ne (e$-$4)                          &1.87  &1.76   &1.90   &1.76   &177    &2.44\\
       Mg (e$-$5)                          &      &       &       &3.8    &12.1   &3.83\\
       Si (e$-$5)                          &      &       &       &3.5    &11.3   &3.53\\
       S  (e$-$5)                          &1.75  &1.75   &1.75   &1.75   &5.16   &1.76\\
       Cl (e$-$7)                          &3.0   &2.56   &2.53   &2.35   &10.1   &2.38\\
       Ar (e$-$6)                          &3.0   &2.75   &3.0    &2.9    &11.5   &2.93\\
       Fe (e$-$6)                          &      &       &       &1.5    &110    &1.92\\
    $\tau_{\rm x}$$^{\rm a}$~(\hi)              &2.61 &8.91  &362.6 &       &       &473\\ 
    $\tau_{\rm z}$$^{\rm b}$~(\hi)              &2.63 &3.69  &5.275  &       &       &6.1\\
    $\tau_{\rm x}$$^{\rm c}$~(\hei)              &0.54 &1.83  &73.72  &       &       &95.7\\
    $\tau_{\rm z}$$^{\rm d}$~(\hei)              &0.54 &0.76  &1.088  &       &       &1.24\\
    $\tau_{\rm x}$$^{\rm e}$~(\heii)              &74.5 &89.1  &120.5 &       &       &90.1\\
    $\tau_{\rm z}$$^{\rm f}$~(\heii)              &77.6 &53.7  &59.93  &       &       &42.1\\
    $\langle$${\it T}_{\rm e}$$\rangle$ (K)            &8,663  &8,825   &8,583   &9,007   &815    &8,892 \\
    \tsq$^{\rm g}$                                   &0.008 &0.006 &0.006 &       &       &0.014\\
     \hline
 \end{tabular}
\flushleft
$^{\rm a}$ Optical depth along the x axis at the \hi~ photoionization threshold.\\
$^{\rm b}$ Optical depth along the z axis at the \hi~ photoionization threshold.\\
$^{\rm c}$ Optical depth along the x axis at the \hei~ photoionization threshold.\\
$^{\rm d}$ Optical depth along the z axis at the \hei~ photoionization threshold.\\
$^{\rm e}$ Optical depth along the x axis at the \heii~ photoionization threshold.\\
$^{\rm f}$ Optical depth along the z axis at the \heii~ photoionization threshold.\\
$^{\rm g}$ As definded by Peimbert 1967\\
\end{table}

\subsection{Updates to {\sc mocassin}}
\label{updates}

For the purpose of our study, we have made a number of updates to {\sc mocassin} Version 2.02.12, including incorporation of 
\hei~line emissivities valid at low temperatures, di-electronic recombination of the third-row elements of the periodic table,
recombination contributions to CELs, and photoionization depopulation of the  He~{\sc i} meta-stable level 2\,$^3$S.

\subsubsection{\hei~ line emissivities}
The original code uses \hei~ line emissivities at one electron density of 10$^2$ cm$^{-3}$ for
\elt~range 5,000--20,000 K, insufficient to treat emission from the cold component ($T_{\rm e}\la 1,000~\rm{K}$) 
present in our bi-abundance model. In addition, emissivities of some He~{\sc i} lines, such as the $\lambda$10830, are 
sensitive to electron density. {\bf Using the calculations of Benjamin et al. (1999) and Smits (1996), 
we have extended the  \hei~ line emissivities in the code to cover  the density range 10$^{2}$--10$^{6}$ cm$^{-3}$ 
and the temperature range 312.5--5,000 K}. Note that emissivities of a few weak He~{\sc i} lines ($\lambda$2946, 
$\lambda$3615, $\lambda$4122, $\lambda$4439, $\lambda$5049, and $\lambda$9466) for this low temperature range are unavailable. 
Collisional excitation from the \hei~2\,$^3$S and 2\,$^1$S meta-stable levels is insignificant at low temperatures, 
and is only considered for \elt~ $>$ 5,000\,K.
{\bf Using the atomic data of Porter et al. (2005), Porter et al. (2007) have provided the most recent \hei~ line emissivities.
Aver et al. (2010) have compared the results of Porter et al. (2007) and Benjamin et al. (1999). 
The differences therein are generally small compared to the ``departure ratios'' in Table~1.}

\subsubsection{Radiative transfer of \hei}
For the \hei~singlets, Case\,B recombination is assumed.  For the \hei~triplets,
the lowest term, 2\,$^3$S, is meta-stable and thus considerably  populated, leading to
significant self-absorption and collisional excitation from this level. 
The \hei~2s\,$^3$S -- 3p\,$^3$P $\lambda$3888 line suffers the strongest suppression due to self-absorption. 
A fraction of those photons, depending on the optical depth, are absorbed and converted into the $\lambda$7065 line photons. 
The \hei~$\lambda$10830 line is strongly affected by collisional excitation. In fact, it is mainly excited by the collisional
process  \hei~2s\,$^3$S -- 2p\,$^3$P. 
Thus, an accurate calculation of the population of level 2s\,$^3$S  is essential to predict reliable fluxes of 
the \hei~$\lambda$$\lambda$3888,7065,10830 lines.
The 2s\,$^3$S level is populated by recombinations of He$^+$ with electrons to He$^0$
triplet levels, and is depopulated by the following processes (Clegg and Harrington 1989): 
(a) collisional excitation to He$^0$ singlet levels; (b) radiative 
decay to the ground state; (c) collisional ionization by thermal electron impacts; (d) photoionization by
UV photons with $\lambda \la$ 2600 {\AA}, such as the resonant \hi~Ly$\alpha$ photons;  
(e) decay from the He~{\sc i} 2p\,$^3$P state to the ground state following 
collisional excitation and the trapping of $\lambda$10830 photons in the nebula. 
Among these  processes, (e) contributes only a few per cent
of the destruction of 2s\,$^3$S and thus can be neglected. Processes (a)--(c) have been considered by 
Benjamin et al. (1999) but not process (d),  because its effects depend on the surrounding radiation 
fields that are unknown without detailed modelling. The original {\sc mocassin}
code does not consider the destruction of meta-stable helium by the photoionization of trapped resonant \hi~Ly$\alpha$ photons,
a process that has been shown to be important for compact and optically-thick PNe (Clegg \& Harrington 1989).
This process is introduced in our updated version. With the He~{\sc i} 2s\,$^3$S level population determined, the
self-absorption of the He~{\sc i} $\lambda$3888 line can be treated using a Monte Carlo approach.
Following the absorption of  a He~{\sc i} $\lambda$3888 photon from the 2s\,$^3$S state to 3p\,$^3$P state, it can either 
scatter back or cascade down through the 3s\,$^3$S and 2p\,$^3$P states with a probability about 0.9 and 0.1, respectively. In the latter case,
three photons ($\lambda$4.3\,$\mu$m, $\lambda$$\lambda$7065,10830) are emitted. The self-absorption problem of He~{\sc i} triplets has
been investigated by Robbins (1968)  by solving an integral equation of transfer for the ideal case of a uniform 
sphere expanding with a constant velocity gradient. For this ideal case, the results deduced from our Monte Carlo approach 
agree within 10 per cent with those of Robbins for $\tau(\lambda3888)$ $\la 10$.

\subsubsection{Di-electronic recombination of S, Cl and Ar}

While no reliable di-electronic recombination rates for the third-row elements have been
published to date, they are estimated to be larger than or at least comparable in magnitude to their
radiative counterparts. Considering the fact that the rates seem to follow a certain trend with ionization stage,
Ali et al. (1991) suggested that for a given ion X$^{i+}$, better estimates of
the di-electronic recombination rates than zero can be obtained by taking the average of the rates
for the second-row elements C$^{i+}$, N$^{i+}$, O$^{i+}$.
Dudziak et al. (2000) adopted coefficients empirically calibrated with
an unpublished model of the PN NGC\,7027. More recently, Ercolano et al. (2004) obtained upper limits
to the rate coefficients of S$^{2+}$, Cl$^{2+}$ and Ar$^{2+}$ by modelling the PN NGC\,1501.
In our modelling, we have assumed that the di-electronic recombination rates of S, Cl and Ar are all
proportional to those of O, and determined their actual values by optimizing the model fit to observations
(see Section~4.1).

\subsubsection{Recombination contributions to CELs}
Radiative cascades following recombination (radiative and di-electronic) of heavy element ions
can partially contribute to the emission of CELs, especially for lines from the meta-stable levels of low-ionization ions.
This process is included when calculating the emissivities of CELs from [C~{\sc i}], [C~{\sc iii}], [N~{\sc i}],
[N~{\sc ii}], [N~{\sc iv}], [O~{\sc ii}], [O~{\sc iii}], [Ne~{\sc iv}] and [Ne~{\sc v}].
Here the relevant radiative and di-electronic recombination rates are taken from P\'equignot et al. (1991) and
Nussbaumer and Storey (1984), respectively, except [N~{\sc ii}] and [O~{\sc ii}], for which the total recombination
rates calculated by  P. J. Storey (private communication) are used.

\subsubsection{Convergence criteria}
{\sc mocassin} uses the Monte Carlo method to simulate the propagation of photons inside a nebula. 
Thus, one of the key component of the code is its convergence criteria. 
The original convergence indicator used in the code is the neutral hydrogen fraction, namely 
a model is deemed as converged if the fraction of neutral hydrogen in most cells
varies less than 5 per cent for two adjacent iterations.
However, only using the fraction of neutral hydrogen as convergence criteria cannot guarantee the 
convergence for other ions, especially for cells in the X$^{i+}$/X$^{(i-1)+}$ transition regions.
For example, we find that in a ``converged'' model based on the original criterion,
the integrated He$^{2+}$/He ratio may vary 10 per cent after additional iterations. 
To overcome this, we have also expanded the convergence criteria to include integrated 
intensities of emission lines from helium and heavy element ions, such as the \heii~$\lambda4686$ and \foiv~25.9\,$\mu$m, 
in our modelling.

\subsection{Central star}

The properties of the central star are basic input parameters in the models.
Based on the detection of broad O~{\sc vi} $\lambda$3811 and C~{\sc iv} $\lambda$5801
features, L2000 classified the central star of NGC\,6153 as a [WC]-PG~1159 H-deficient star.
The classification is supported by a study of the UV spectrum of the central star (Pottasch et al. 2003).
Using the blackbody approximation, we have calculated the effective
temperature (\teff), luminosity, core mass, and surface gravity of the central star using the Zanstra method (Zanstra 1931) and 
the Stoy method (Stoy 1933). The results are listed in Table~3. In the calculations, we have assumed the distance to 
NGC\,6153 to be 1.5\,kpc (P\'equignot et al. 2003).  
Both the Zanstra and Stoy methods require a measurement of 
the stellar continuum flux of the central star. For this purpose we have made use of the {\it HST}/WFPC2 F502N image and find 
a stellar continuum flux of $2.75\times10^{-14}$ ergs~cm$^{-2}$~s$^{-1}$~{\AA}$^{-1}$ at 5\,010\,{\AA} after subtracting 
the nebular background emission.
The applicability of the Zanstra method also requires the nebula to be optically thick, which is 
probably not the case for the \hi~Lyman continuum given that NGC\,6153 is a high-excitation class PN and 
is probably matter bound. 
Thus the Zanstra temperature deduced from the H$\beta$ flux gives only a lower limit, while that
deduced from the \heii~$\lambda$4686 line is closer to the actual value
since the nebula is likely to be optically thick in the \heii~Lyman continuum.
The Stoy method has the advantage that it does not depend on the nebular optical depth. However,
the applicability of this method requires the measurement of the total flux of all cooling lines from the UV to the far-IR, not an 
easily achievable task. 
We find that the Stoy temperature is lower than the He~{\sc ii} Zanstra
temperature, suggesting that the total flux of the nebular cooling  emission lines
might have been underestimated.
\setcounter{table}{2}
\begin{table} \centering \caption{Parameters of the central star.}
\label{central star} 
\begin{tabular}{|l|l|l|l|l|} \hline
 &\teff  &  $L$         & ${\it M}_{\rm core}$  & $\log g $  \\
 &(${\rm  K}$)            & (${\rm L}_{\odot}$) & (${\rm M}_{\odot}$)   &  (${\rm {cm~s^{-2}}}$) \\\hline
Stoy method               &76,530&2,034&0.562&5.37  \\
H {\sc i} Zanstra method  &76,260&2,014&0.562&5.37  \\
He {\sc ii} Zanstra method&90,090&3,228&0.586&5.47  \\\hline
\end{tabular}
\end{table}

\subsection{Spherical model}
\label{sphmodel}

\subsubsection{Model parameters}

An initially chemically homogeneous spherical model (referred as model~S hereafter) is constructed.
Its density distribution is constrained by the azimuthally averaged radial surface brightness distribution of 
\ha~deduced from the {\it HST}/WFPC2 F656N image.
We find that a density distribution given by
\begin{equation}\label{formula1}
    \rho_s(r,\theta,\Phi)=N_0(s{\times}f(r|a_1,b_1)+(1-s){\times}f(r|a_2,b_2))
\end{equation}
where
\begin{eqnarray}\label{formula2}
  f(r|a,b)=\frac{1}{4\pi b^a\Gamma(a)}r^{a-3}\exp(-r/b)
\end{eqnarray}
can reasonably reproduce the observed H$\alpha$ surface brightness distribution, as shown in Fig.~2. The parameters are: 
$a_1=23$, $b_1=0.0118$, $a_2=12$, $b_2=0.037$, and $s=0.2$. Note that the shell is actually composed of two density profiles
with peaks at $r~=~(a_{1,2}-3)\times b_{1,2}$ and FWHMs $(a_{1,2}-3)/b_{1,2}$, as illustrated in Fig.~3. In Eqs~(1)
and (2), $r$ ranges from 0 to 0.5, corresponding to a nebular angular radius of $0\arcsec$ to $16\arcsec$.
A black body spectral energy distribution for the ionized source is assumed.  The model was constructed taking the 
\teff~and luminosity of the central star, the  nebular elemental abundances and the total mass as
free parameters.

\begin{figure}
\label{sphfluxdist}  
  \centering
  \epsfig{file=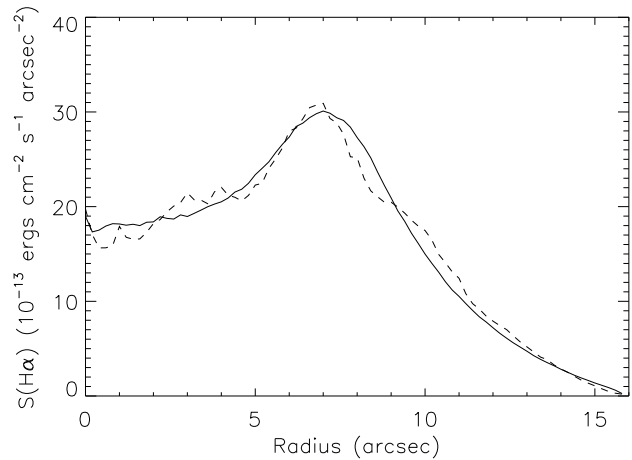, width=8.5cm, angle=0}

  \caption{Comparison between the observed (dashed line) and model-predicted (solid line) azimuthally averaged radial surface brightness 
distribution of \ha~{\bf of model S}.}
\end{figure}

\begin{figure}
\label{sphdensitydist}  
  \centering
  \epsfig{file=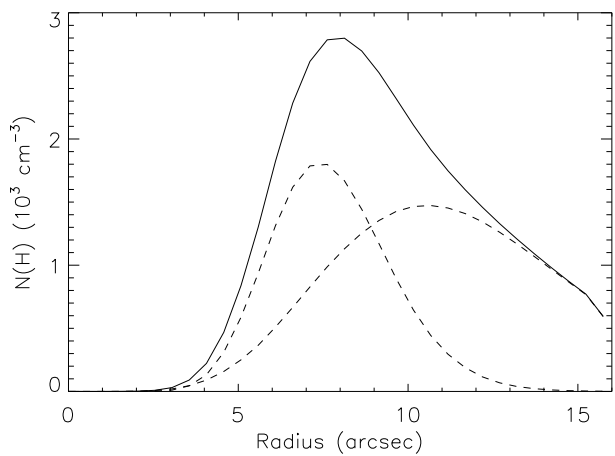, width=8.5cm,angle=0}

  \caption{Radial number density distribution of hydrogen atoms N(H) of the spherical model (solid line). 
It is composed of two components with different profiles (dashed lines).}
\end{figure}

\subsubsection{Model results}
The spherical model has a  total mass of 0.35 M$_{\odot}$ and is
fully ionized by the central star which has a \teff~ of 90,000 K and a luminosity of
4,000 L$_{\odot}$. The nebula is optically thin in both the \hi~and \hei~ionizing continua,
but is optically thick in the \heii~ionizing continuum.
The electron temperature throughout the nebula is remarkably uniform.
We obtain a mean electron temperature of 8,663\,K, lower than the value derived
from the \foiii~temperature diagnostic lines, and a small mean-square temperature fluctuation parameter 
$t^2$ (as defined by Peimbert 1967) of 0.008.

Fig.~4 shows the radial temperature and
density distributions and the ionization structures of the spherical model. 
In Fig.~4 (also in Figs. 7 and particularly 13), the sharp temperature dip in the high-ionization region 
($r \sim 3-4\arcsec$) is due to efficient cooling by the  strong \fnev~ fine-structure lines at 24.2\,$\mu$m and 14.3\,$\mu$m. 
This feature, which is a consequence of the low (but not zero) density assumed close to the star in the present model, 
has negligible effect on the model nebular emission.
The decline of temperature for $r$ between $4\arcsec$ and $8\arcsec$ is caused by the enhanced cooling of the \foiii~lines,
as the ionic concentration of O$^{2+}$ increases with $r$.

The predicted over observed line flux ratios (hereafter  'departure ratios')  are
given in the 4th column of Table~1. Most hydrogen and helium recombination lines
are well matched by the model. The small remaining discrepancies between observed
and predicted fluxes are within observational uncertainties, except
for the \hei~$\lambda$$\lambda$7281,10830 lines. The latter two lines with large
departure ratios will be further discussed in Section~4. The model however fails to explain 
the strengths of all heavy element ORLs where the model predictions are generally
about one order of magnitude lower than the observed values.

The fluxes of CELs from high-ionization species are reasonably reproduced by the model.
For low-ionization species, such as \fci, \fni, \foi, \foii~and \fsii, the predicted
fluxes are significantly lower than observed. This can be ascribed to the oversimplified
density distribution assumed in the spherical model. The nebula is clearly ellipsoidal and bipolar, 
and appears to be ionization-bound at least in some directions,
where low-ionization species can survive. We infer that there exists a dense torus and present two
ellipsoidal models without and with a torus in the next subsection.

\begin{figure}
\label{sphtenehhe}  
  \centering
  \epsfig{file=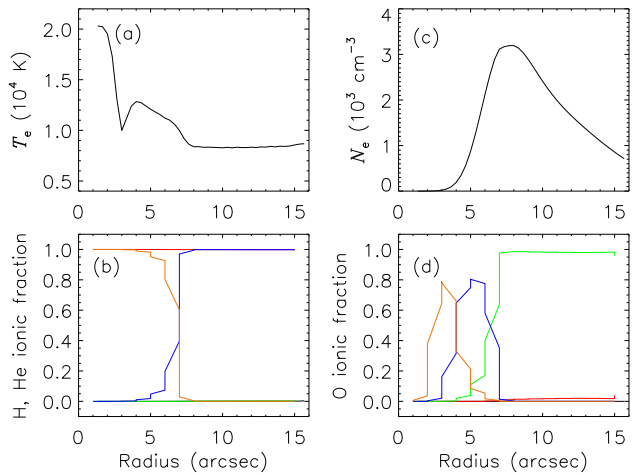, width=8.5cm, angle=0}
  \caption{Distributions of \elt~(panel a); \eld~(panel c); H, He (panel b) and O (panel d) ionization structures of the spherical model.
  In panel b, the yellow, blue, green, red, and black lines represent those of
  He$^{2+}$, He$^{+}$, He$^{0}$, H$^{+}$ and H$^0$, respectively. In panel d, the yellow, blue, green, red, and black
  lines  represent those of O$^{4+}$, O$^{3+}$, O$^{2+}$, O$^{+}$ and  O$^0$, respectively.}
\end{figure}

\begin{figure*}
\label{eb}  
  \centering
  \epsfig{file=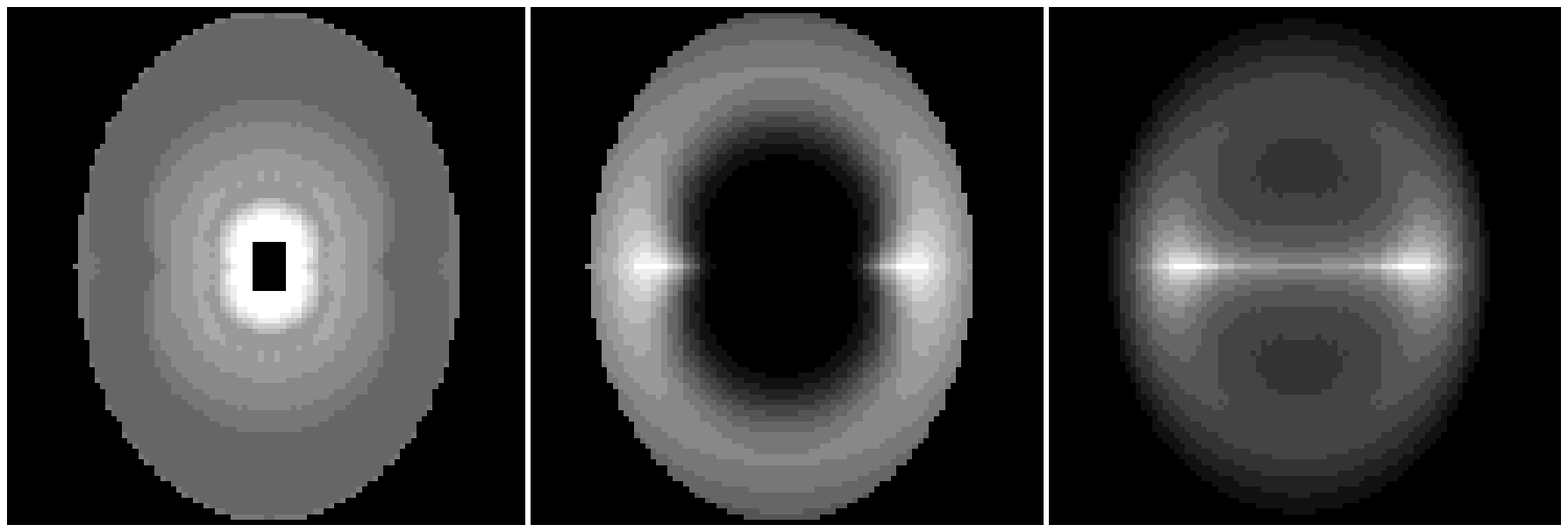, width=18cm,angle=0}
  \caption{Distributions of \elt~ (left panel), N(H) (middle panel) and the surface brightness of 
\ha~(right panel) in the x-z plane for model E1. The contours are of arbitrary units.}
\end{figure*}

\begin{figure*}
\label{ebtdensity}  
  \centering
  \epsfig{file=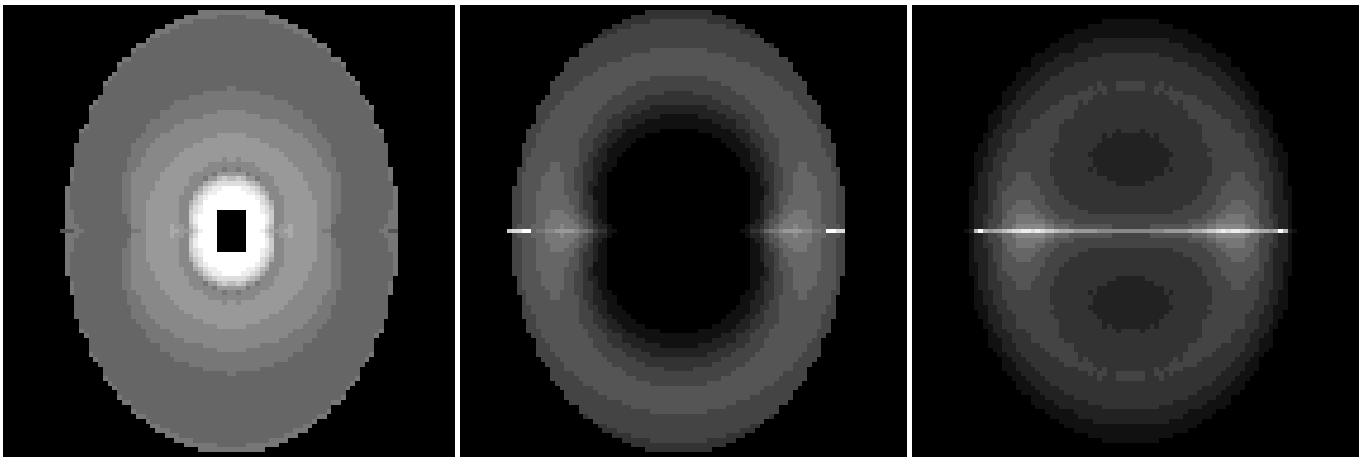, width=18cm,angle=0}
  \caption{Distributions of \elt~ (left panel), N(H) (middle panel) and the surface brightness of 
\ha~(right panel) in the x-z plane for model E2. The contours are of arbitrary units.}
\end{figure*}

\subsection{Ellipsoidal models}

\subsubsection{Model parameters}

To investigate possible effects of the geometry, we have built two  ellipsoidal models without (model E1)
and with (model E2) a torus. Based on the {\it HST} images, we assume that the nebula has a semimajor axis of 16$\arcsec$ in 
the polar (z-) direction and a semiminor axis of 12$\arcsec$ in the equatorial (x- and y-) plane. The density distribution
of the ellipsoid is set by
\begin{equation}\label{formula3}
    \rho_e(r,\theta,\Phi)=\frac{\rho_s(\frac{r}{1+c|\cos(\theta)|})}{1+d|\cos(\theta)|^g},
\end{equation}
where $\rho_s$ was defined in Eq~(1). For the ellipsoidal models, the parameters $a$, $b$, and $s$
are set to the same values as those adopted in model S. We introduce three new parameters $c$, $d$, and $g$ to characterize
the density gradient in the equatorial plane and the polar direction. The values of $c$, $d$, and $g$
are constrained by the {\it HST} images as well as the observed fluxes of low- and high- excitation CELs. 
Via model fitting, we obtain $(c,d,g)=(1,0.8,0.5)$  for model E1 and $(0.8,0.5,0.5)$ for model E2. A torus
with a width of 6 cells and a height of 1 cell is added in model E2. Its density is assumed to be 2.5 times
higher than the original values of the replaced cells. The  density distributions of model~E1 and model~E2
in the x-z plane are displayed in the middle panels of Figs.~5 and 6, respectively.

\subsubsection{Model results}

Figs.~5 and 6 display the distributions of the predicted temperature, hydrogen number density and 
the surface brightness of H$\alpha$ of models E1 and E2, respectively. 
Fig.~7 shows the model radial distributions of the electron temperature, density
and ionization structures. In Fig.~8, we compare the  predicted
line fluxes of model E2 with observations. We find that 
both models E1 and E2 give a better fit to the \fnii, 
\foii, and \fsii~ lines compared to model S thanks to the density enhancement in the equatorial plane.  
Model E2 yields a better fit for the \fci~and \fni~ lines than model E1, owing to the presence of a dense torus. 
For the remaining lines, models E1 and E2 give similar results of model S.
Although the ellipsoidal models show improvements in reproducing CELs from
low-ionization ions, like model S, they fail to reproduce the strengths of heavy element ORLs, 
strongly indicating that chemical inhomogeneities have to be invoked.

\begin{figure*}
\label{eb_structure}
\centering
\epsfig{file=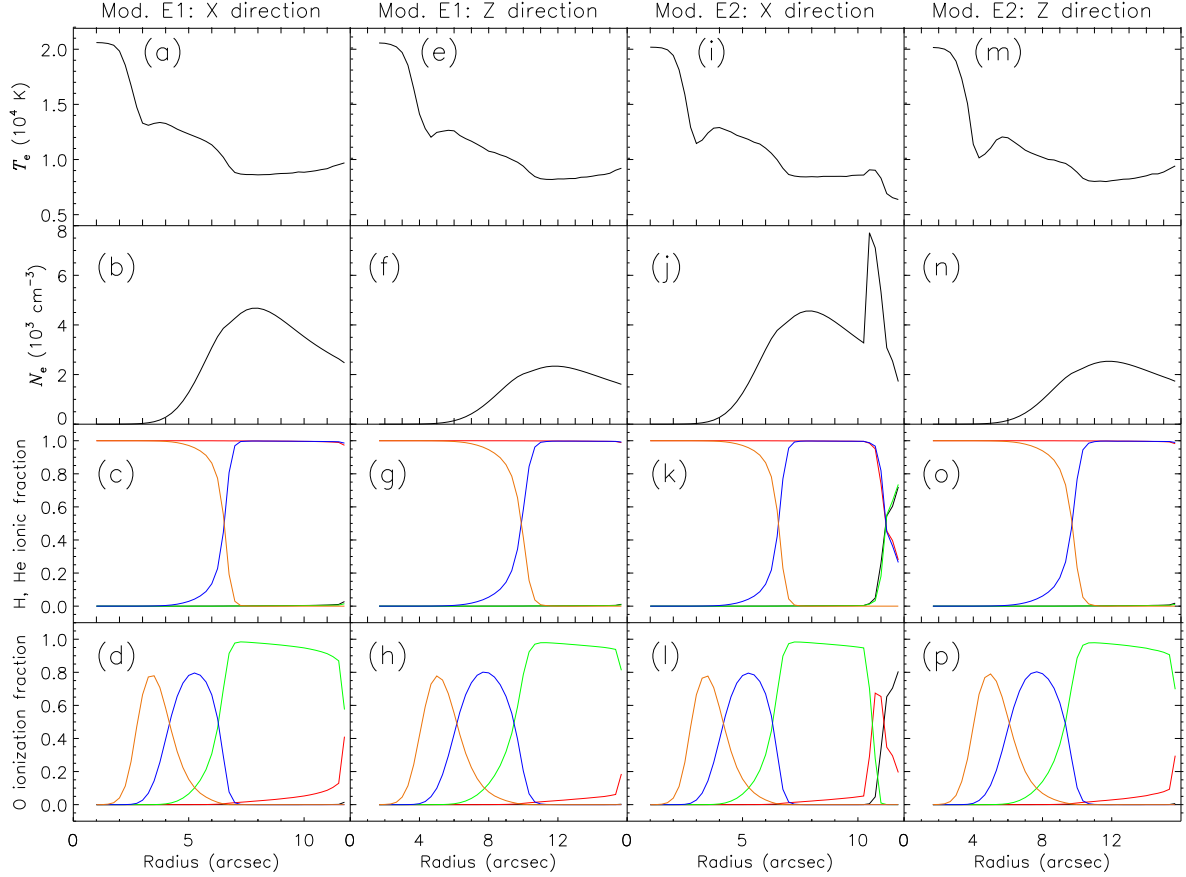, width=16cm}
\caption{\elt~and \eld~ distributions and H, He and O ionization structures in the two ellipsoidal models.
Panels (a) -- (d) and (e) -- (h) are for the x, z directions in the ellipsoidal model E1, respectively;
Panels (i) -- (l) and (m) -- (p) are for the x, z directions in the ellipsoidal model E2, respectively.
In panels (c), (g), (k), and (o), the yellow, blue, green, red and black lines  represent the fractions of 
He$^{2+}$, He$^{+}$, He$^{0}$, H$^{+}$ and H$^0$, respectively. In panels (d), (h), (l), and (p), the yellow, blue, green, red, and black 
lines represent the fractions of O$^{4+}$, O$^{3+}$, O$^{2+}$, O$^{+}$ and  O$^0$, respectively. Note that the sharp
features near 10\farcs5 in panels (i), (j), (k) and (l) are produced by the dense torus.
}
\end{figure*} 

\begin{figure*}
\label{ebtbarplot}  
  \centering
  \epsfig{file=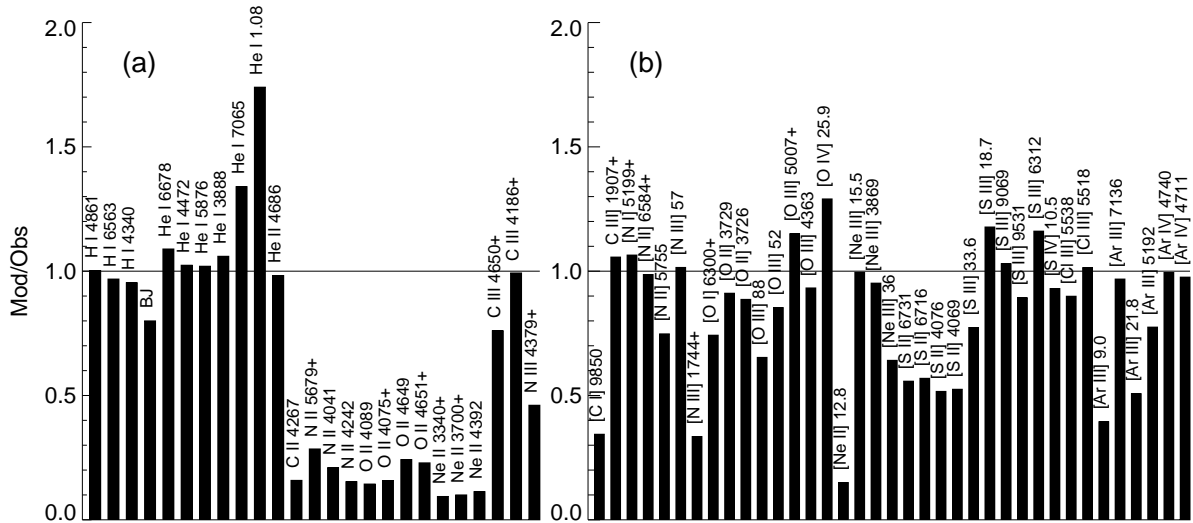, width=16cm,angle=0}
 \caption{ Comparison of line fluxes predicted by model E2 and observations for
ORLs (left panel) and CELs (right panel). }
\end{figure*}

\subsection{Bi-abundance model}

\subsubsection{Model parameters}

For the bi-abundance model, we assume that there are some metal-rich knots (cells) embedded  in the
diffuse nebula of ``normal'' abundances. The density distribution of the diffuse nebula is
inherited from model E2, but with a slightly reduced mass. 
The spatial distribution of the metal-rich cells is set randomly
by the probability density function $ f(r|a_3,b_3)$ (see Eq~(2)), where $r$ ranges from 0 to 0.3125 (i.e. 0--10\arcsec), $a_3=22$ and $b_3=0.013$,
as constrained by the surface brightness distribution of the \oii~$\lambda$4649 ORL 
along the nebular minor axis (see Fig.~19 of L2000). 
In Fig.~9, we compare the profile of the  predicted \oii~ surface brightness with observations.
In the comparison, we have included the effects of seeing and convolved the predicted surface brightness distribution with a 
Gaussian function of FWHM $3\farcs2$.   
Positions of individual knots are set by Monte Carlo simulations. Their three-dimensional spatial distribution is illustrated in Fig.~10. 
Fig.~11 shows the distributions of number density of knots  and of the hydrogen atoms of the diffuse gas.
In order to reduce the random uncertainties, we have generated four samples of knots. 
Line fluxes yielded by the four samples are in good agreement (within 10\,per cent). They are then averaged to 
compare with the observations. The model parameters are given in Table~2, where B$_{\rm n}$ and B$_{\rm c}$ represent
respectively components of ``normal'' composition  and of cold metal-rich knots, and B is the sum of the two
components. Note that for the normal component, we have fixed the helium abundance to 0.1 (as adopted by P\'equignot et al. 2003).
As pointed out by P\'equignot et al., observations of helium recombination lines in the optical alone are not sufficient 
to decompose the helium abundance in the two components. For the H-deficient knots, the element abundances heavier than neon 
are assumed to be the corresponding solar values in mass, considering that PNe originate from low- and intermediate- mass stars
that can not process the third-row elements.

\begin{figure}
\label{plotoii}  
  \centering
  \epsfig{file=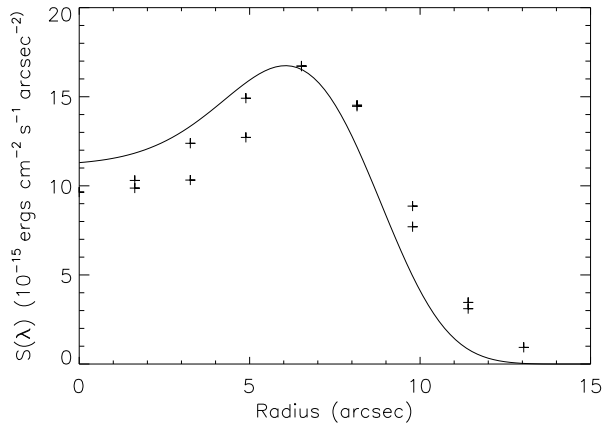, width=8cm,angle=0}
  \caption{ Comparison of the measured surface brightness distribution (``+'') of the \oii~$\lambda$4649 ORL and the
model prediction (solid line).}
\end{figure}

\begin{figure}
\label{cells_3D}  
  \centering
  \epsfig{file=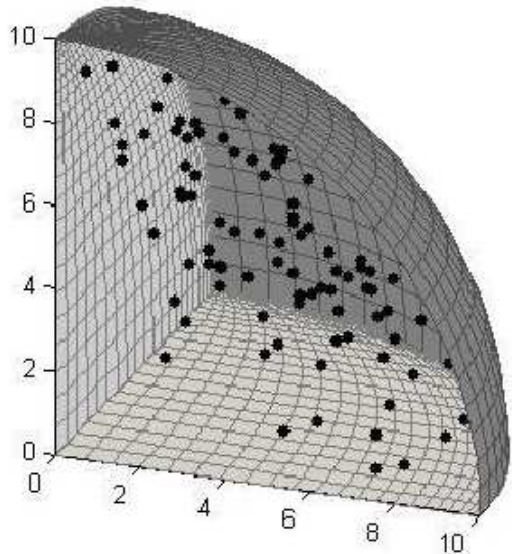, width=8cm,angle=0}
  \caption{Three-dimensional spatial distribution of metal-rich knots (cells) in the bi-abundance model.
The units of the axes are arcsec.}
\end{figure}

\begin{figure}
\centering
\epsfig{file=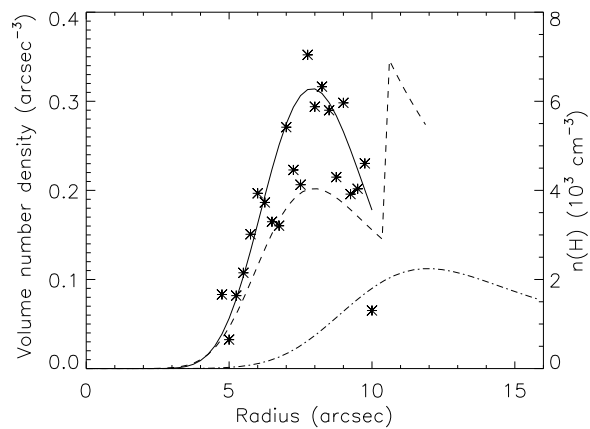, width=8cm,angle=0}
\caption{Radial distribution of the volume number density of metal-rich knots given by the analytic 
(solid line) and Monte Carlo simulations (asterisks). The dashed and dash-dotted lines represent the density distribution
of hydrogen atoms in the diffuse nebular gas along the x and x axes, respectively.}
\end{figure}

\subsubsection{Model results}

Fig.~12 displays the predicted structure of \elt~and \eld. The radial distributions of
temperature, density and the ionization structures in four directions across
the four metal-rich knots marked in Fig.~12 are shown in Fig.~13.
An inspection of the two figures shows that the knots have been cooled down to very low temperatures ($\sim800$\,K),
and create shadows that block the ionizing UV radiation field from the central star, leaving low ionization regions behind them.
As in models S, E1 and E2, there is a temperature dip at $r\sim 3-5\arcsec$. 
Fig.~14 displays the predicted monochromatic images of H$\beta$, [O~{\sc iii}]~$\lambda$5007, and O~{\sc ii}~$\lambda$4649.
The figure clearly shows that the [O~{\sc iii}]  CEL originates from the ``normal'' component, while the O~{\sc ii}
ORL is dominated by knots. In spite of their small number, metal-rich knots have a non-negligible  
contribution to H$\beta$ emission, owing to the enhanced emissivity at low temperatures and high densities.
In Fig.~15, we compare the predicted line fluxes with the observations, demonstrating that the
bi-abundance model has significant improvements in reproducing the strengths of heavy element ORLs and the hydrogen Balmer discontinuity. 
In addition, the model yields a much better fit to the measured flux of the \fneii~12.8\,$\mu$m line, which is underestimated
by all the chemically homogeneous models. This is due to the fact that the metal-rich knots are enriched in Ne and have a 
large Ne$^+$ ionic concentration.

{\bf Model B generally supports the results of the one-dimensional models of P\'{e}quignot et al. (2002, 2003) 
in terms of properties of the H-deficient inclusions such as the electron temperature, density, mass and chemical abundances.
Those values in model B (\elt$~$= 1,390 K, ${\it N}({\rm H})$ = 4,410 \cmt,  $M$ = 0.0031$\mathrm{{\it M}_\odot}$ 
and oxygen enrichment factor of about 100) are comparable to those given by the one-dimensional models  
(\elt$~$= 815 K, ${\it N}({\rm H})$ = 4,000 \cmt,  $M$ = 0.0031$\mathrm{{\it M}_\odot}$ and oxygen enrichment factor of about 80).
Beyond constraining the properties of the H-deficient inclusions in a self-consistent manner, the three-dimensional model 
enables to investigate sizes and spatial distributions of the H-deficient inclusions which one-dimensional models 
can not. Detailed line by line comparison between model B and the one-dimensional models by P\'{e}quignot et al. (2002, 2003) 
is not easy, due to different geometric configurations, density structures, spectral energy distributions of 
the ionizing star and atomic data used. We do not claim superiority of predicted fluxes of model B over the one-dimensional models.
To understand how three-dimensional bi-abundance models advance the field of modelling of emission line 
gaseous nebulae, simple bi-abundance benchmark models are needed. Then the comparison between three-dimensional model results 
and one-dimensional model results will be much easier to carry out and interpret, but this is beyond the scope of the current paper.
}
\begin{figure*}
\centering
\epsfig{file=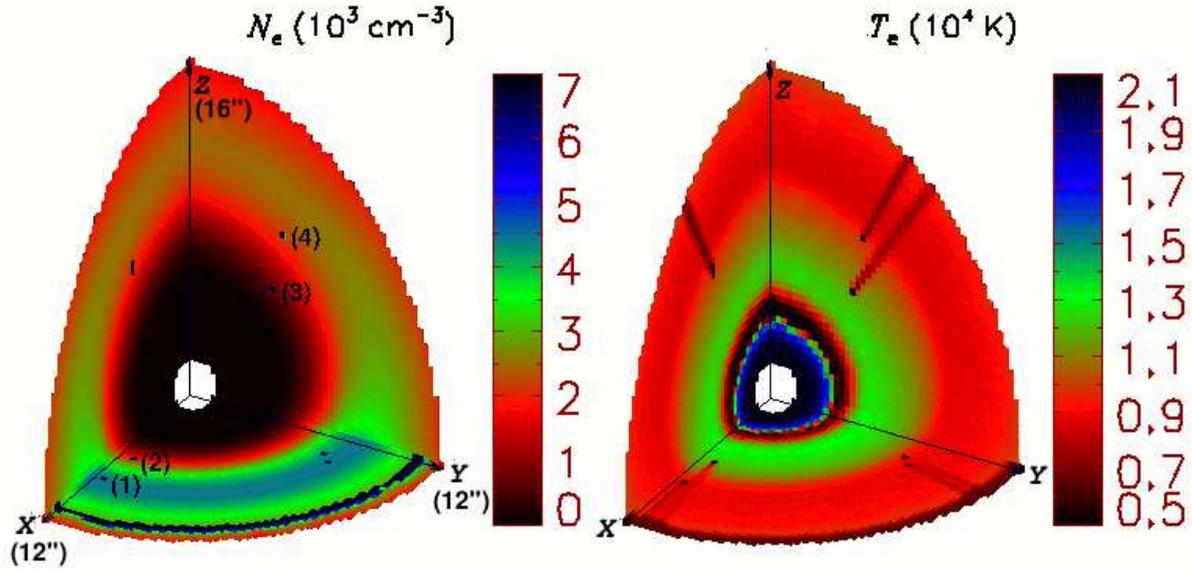, width=16cm}
\caption{Predicted \eld (left panel) and \elt~(right panel) distributions in the bi-abundance model B.}
\end{figure*}

\begin{figure*}
\label{bi_structure}
\centering
\epsfig{file=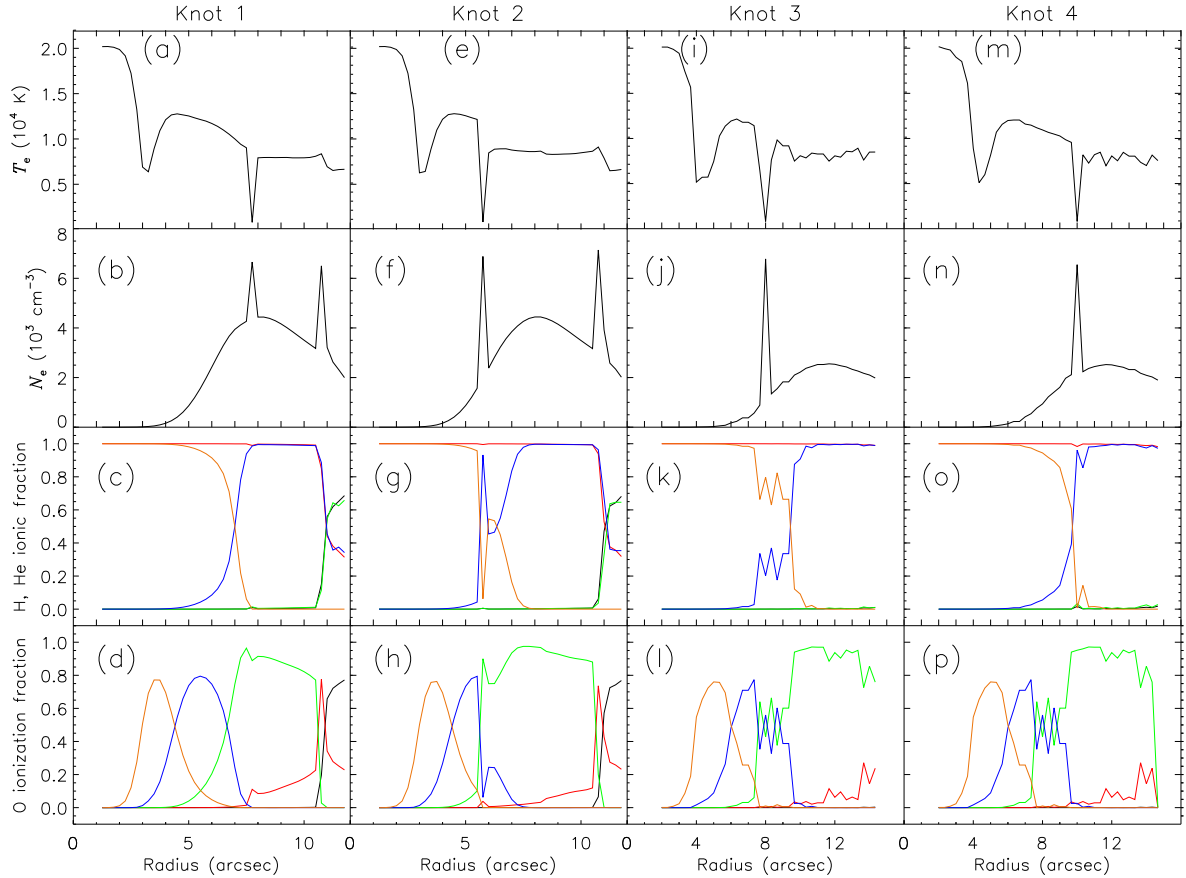, width=16cm}
\caption{Distributions of \elt, \eld~ and H, He and O ionization structures along a radial direction that passes through
the marked four knots in Fig.~12.  Panels (a) -- (d) and (e) -- (h) are for the Knots 1 and 2 located in the xy plane, respectively;
Panels (i) -- (l) and (m) -- (p) are for the Knots 3 and 4 in the yz plane, respectively.
In panels (c), (g), (k) and (o), the yellow, blue, green, red, and black lines represent the fractions of 
He$^{2+}$, He$^{+}$, He$^{0}$, H$^{+}$ and H$^0$, respectively. In panels (d), (h), (l) and (p), the yellow, blue, green, red, and black 
lines represent the fractions of O$^{4+}$, O$^{3+}$, O$^{2+}$, O$^{+}$ and  O$^0$, respectively.
Note that in panels (b) and (f), the peaks at $\sim8\arcsec$ and $\sim11\arcsec$ correspond respectively to the knot and the torus.
}
\end{figure*}

\begin{figure*}
\centering
\epsfig{file=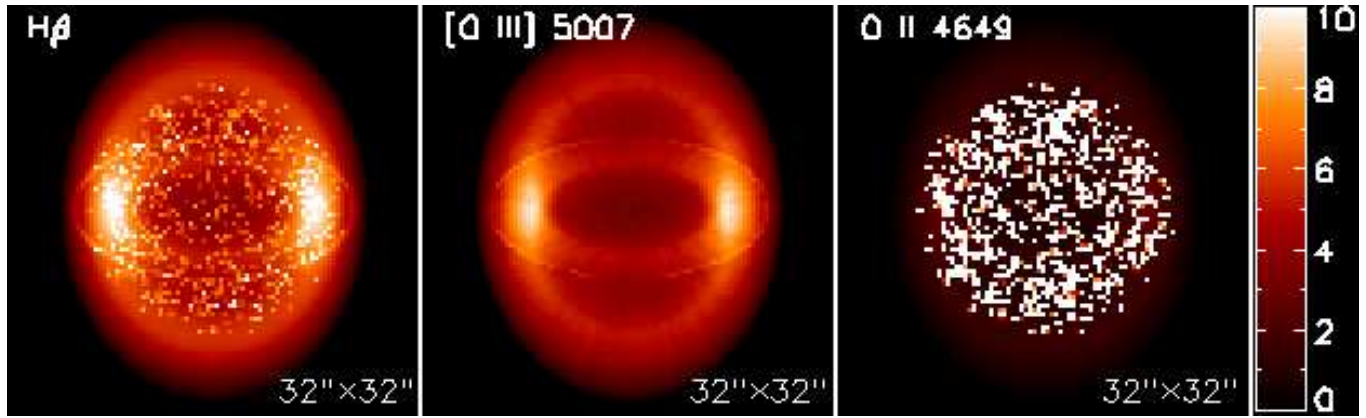 ,width=18cm}
\caption{Projected monochromatic images of \hb~(left panel), \foiii~$\lambda$5007 (central panel) and O~{\sc ii}~$\lambda$4649 (right panel)
predicted by the bi-abundance model B at a viewing angel of $\theta$ = 60 deg. $\theta$ = 0 deg is for face on.
The units of the color bars are $10^{-13}, 10^{-12} ~{\rm and}~ 10^{-15}~ {\rm ergs}~ {\rm cm}^{-2}~ {\rm s}^{-1}~ {\rm arcsec}^{-2}$
for the three images, respectively.
}
\end{figure*}

\begin{figure*}
\label{bibarplot}  
  \centering
  \epsfig{file=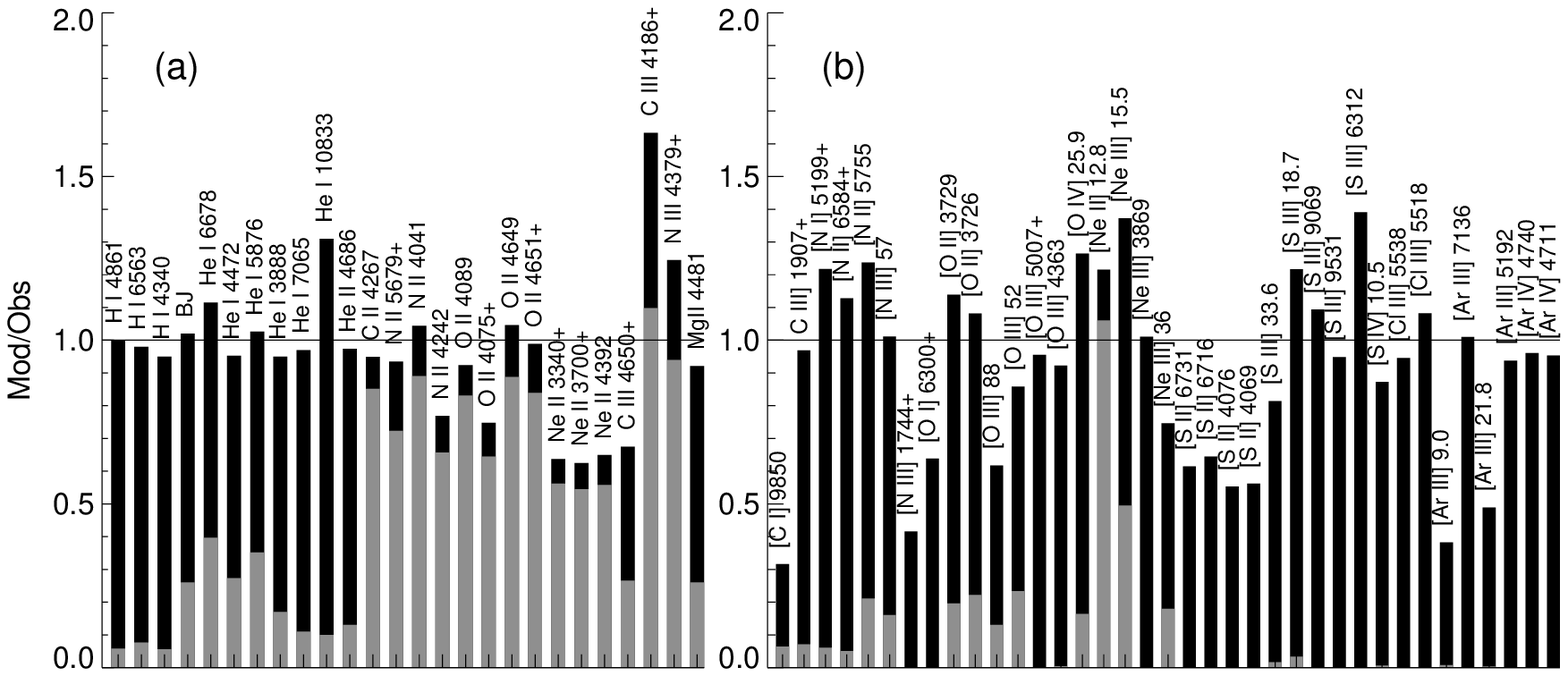, width=16cm,angle=0}
  \caption{Comparison of the predicted and observed line fluxes  for ORLs (left panel) and CELs (right panel) in the
  bi-abundance model B. Black and gray parts represent contributions from the normal and the cold H-deficient components, respectively.}
\end{figure*}

\section{Discussion of the bi-abundance model}
\label{discussion}

\subsection{Uncertainties and discrepancies}
Uncertainties in our models include errors in the atomic data, in the treatment of the radiation 
transfer and uncertainties in the adopted line fluxes. 
The estimates of the di-electronic recombination rates ($\alpha_d$) for S, Cl and Ar are rather rough. 
We have assumed that $\alpha_d$(S, Cl, Ar)~=~$\eta$(S, Cl, Ar)~$\times$~$\alpha_d$(O), where the constants $\eta$(S, Cl, Ar) 
are obtained by fitting the observations.  Only two \fcliii~ lines are detected, insufficient to place a constraint on
the $\eta$(Cl) value. We have thus assumed $\eta$(Cl)~$=1$. S and Ar have a number of lines detected from several ionization stages.
We obtain $\eta$(S, Ar)~=~(0.76, 6.8), (0.45, 4.0), (0.35, 4.0) and (0.35, 4.0) from the models
S, E1, E2 and B, respectively. In the bi-abundance model,  the di-electronic recombination coefficients 
of S and Ar are respectively 0.32 and 14 times the corresponding  radiative recombination values. 
Our estimates of $\alpha_d$(S, Ar) differ from those obtained from the modelling of NGC\,7027 (Dudziak et al. 2000) 
by a factor of three. On the other hand, these uncertainties in the di-electronic recombination rates of S and Ar are 
unlikely to have a large impact on our model predictions for other lines, considering the low abundance of Ar and the 
fact that [S~{\sc ii}], [S~{\sc iii}] and [S~{\sc iv}] lines have similar emissivities, and consequently 
thermal structure is hardly affected by the uncertainties in the ionization 
structure of S.

The bi-abundance model underestimates the flux of C~{\sc iii} multiplet M1, but overestimates that of M18.
This is probably caused by errors in the atomic data. The existing calculations 
of the radiative (P\'equignot et al. 1991) and di-electronic (Nussbaumer \& Storey 1984) recombination coefficients 
for C~{\sc iii} may not be applicable to plasma at very low temperatures.
Further theoretical work is urgently needed to study the behavior of recombination
lines in extremely cold plasma.  We note that because of a larger di-electronic contribution
the emissivity of lines from C~{\sc iii} multiplet M18 is much more 
sensitive to electron temperature than that of M1. Should it not be the case, 
the model would have yielded a much  better fit to both M1 and M18.

The bi-abundance model overestimates the flux of the \hei~$\lambda$10830 line by 21 per cent.
 The \hei~$\lambda$10830 line is crucial for the bi-abundance model in that, unlike other
\hei~ lines, it is predominantly collisionally excited, and thus has the potential to constrain the helium abundance in the 
hot component of ``normal'' composition. Its usage is however hindered by the complexity in the 
\hei~2s$^3$S meta-stable level population and the excitation and radiative transfer of the  
\hei~$\lambda$10830 line. Observations of several PNe have shown that the measured He~{\sc i} line ratio
{\it I}($\lambda$10830)/{\it I}($\lambda$5876) is lower than the theoretical prediction (see, e.g., Peimbert \& Torres-Peimbert 1987a,b).
Possible explanations include the destruction of \hei~$\lambda$10830 photons by internal dust (Kingdon \&
Ferland 1995), and the photoionization depopulation of the \hei~meta-stable level  
by \hi~Ly$\alpha$ photons (Clegg \& Harrington 1989).
The later process has been considered in our model. However its inclusion reduces the predicted flux of the \hei~$\lambda$10830
line only slightly. Additional destruction mechanisms may be needed to reconcile observations with theory.

Given its high sensitivity to electron temperature, the \hei~$\lambda$7281 line serves as an important diagnostic to 
probe the physical conditions in the cold H-deficient knots (Zhang et al. 2005a). 
Although improved compared to the chemically homogeneous models, our bi-abundance model overestimates
the \hei~$\lambda$7281 line by 55 per cent. The discrepancy is probably caused by departure from 
the Case\,B assumption. Under Case\,A recombination, the predicted flux of the \hei~$\lambda$7281 line is 
about a factor of two lower. L2000 and Liu et al. (2001) find that the observed fluxes of the 
2s$^1$S-np$^1$P and 2p$^1$P-ns$^1$S series are systematically lower than predicted
by Case\,B recombination. Similar discrepancies are also found in H~{\sc ii} regions, 
such as the Orion Nebula (Porter et al. 2007; Blagrave et al. 2007) and 30 Doradus (Tsamis \& P\'equignot 2005). 
{\bf Porter et al. (2007) pointed out that although the escaping of the \hei~Lyman photons could cause the discrepancies,
it is not consistent with the fact that the \hei~$\lambda$6678 line flux is not decreased compared to prediction. We also want to point out
that in the case of the escaping \hei~Lyman photons, the discrepancy factors for the np$^1$P-2s$^1$S series ($n=3,4,5$)
will decrease with increasing $n$ values, which is also not seen in the Orion Nebula.
In their non-Case-B, self-consistent ``model III'', Porter et al. (2007) attributed  the discrepancies to ``inaccurate reddening corrections''. 
Another plausible mechanism is that \hei~Lyman photons are absorbed by hydrogen atoms or dust grains, as proposed by Liu et al. (2001).}
We conclude that while Case\,B recombination may still remain a good approximation for the \hei~singlet lines in some PNe
(e.g. NGC\,7027; Zhang et al. 2005b), significant departures from this approximation may occur in others, and the effects
can be quite significant for the 2s$^1$S-np$^1$P and 2p$^1$P-ns$^1$S series.

We deduce the  extinction coefficient $c({\rm H}\beta)$ by comparing the observed \hi~and \heii~ line ratios to 
theoretical values, assuming a constant electron temperature and density.
In the presence of a cold, H-deficient component, such as in the bi-abundance model, the extinction coefficient
thus derived would be overestimated, leading to the fluxes of lines of short wavelengths being systematically overestimated
while those of long wavelengths underestimated. This effect is nevertheless small, less than 5 per cent in the current bi-abundance model.

\subsection{ \elt~ fluctuations versus the bi-abundance model}

\elt~fluctuations have been proposed to explain the CEL/ORL abundance discrepancies as well as
the differences between the  electron temperatures derived from the [O~{\sc iii}] CELs and from the hydrogen Balmer discontinuity
(e.g. Peimbert \& Peimbert, 2006). However, chemically homogeneous models predict
a very small $t^2$.  Using the formula of Peimbert (1967) and the values of $T_{\rm e}$([O~{\sc iii}]) and $T_{\rm e}$([H~{\sc i}]) measured by L2000, 
we find $t^2=0.065$, much larger than the values predicted by chemically
homogeneous models ($t^2 \leq 0.008$; Table~2). 
The bi-abundance model yields a larger value ($t^2 \simeq 0.014$), 
but still significantly lower than the 0.065 derived from the measured $T_{\rm e}$([O~{\sc iii}]) and $T_{\rm e}$([H~{\sc i}]).
Note that the formalism of Peimbert (1967) is only applicable to situations of small amplitudes of temperature variations. 
It will significantly overestimate $t^2$ in the case where the nebula is composed of two distinct components 
of very different temperatures (Zhang et al. 2007).
By satisfactorily reproducing observations of the [O~{\sc iii}] CELs and the hydrogen Balmer discontinuity,
the bi-abundance model provides a natural explanation for the discrepancy between  $T_{\rm e}$([O~{\sc iii}]) and $T_{\rm e}$([H~{\sc i}]).

\subsection{Properties of the cold component}

Table~4 summarizes the properties of the H-deficient knots derived from the bi-abundance model.
The model shows that the knots are fully ionized and optically thin,
with \elt~ of 800 K, a hydrogen number density of 4,000 cm$^{-3}$, a He/H abundance  ratio of 0.5, and CNONe abundances relative 
to hydrogen about 40--100 times higher than the ``normal'' component.
The mass of this cold H-deficient component is only 0.0031~M$_{\odot}$ ($\sim 3$ Jupiter masses), about 1 per cent of the total mass of the nebula.

With much enhanced metal abundances compared to the ``normal'' component, the cooling of the cold component is 
dominated by the IR fine-structure lines from C, N, O and Ne ions. These fine-structure lines have different excitation temperatures,
typically much lower than those of optical/UV CELs. Their emissivities are thus less sensitive to temperature than optical/UV CELs. 
These IR lines also have relatively low, yet varying critical densities. 
Therefore, depending on the local physical conditions, their contributions to the cooling vary. 
It follows that the electron density plays an important role in 
setting the equilibrium temperatures of the cold component by determining which IR lines are dominant coolants. 
At low electron densities,  the \foiii~52 and 88\,$\mu$m lines are the dominant coolants,
whereas at higher densities, the \fneii~12.8\,$\mu$m and \fneiii~15.6, 36\,$\mu$m lines, owing to their higher 
critical densities and higher excitation temperatures than the \foiii~lines, start to dominate the cooling. 
Because the \fneii~ and \fneiii~ IR lines  have higher excitation temperatures than the \foiii~IR lines, the equilibrium temperatures in 
high-density knots are higher than those in low-density ones. For example, as hydrogen number density decreases from 10,000 cm$^{-3}$ 
to 1,000 cm$^{-3}$, the equilibrium temperature drops from 1,070 K to 460 K. It follows that both the density and temperature contrasts between 
the cold component and the normal component are expected to grow as the nebula expands, assuming  that the two components are in 
pressure equilibrium. As a consequence, one expects the adf value to increase as the nebula evolves, consistent with what is
observed (e.g. Zhang et al. 2004). The  equilibrium temperatures in the H-deficient knots are also 
affected by the He and heavy element abundances in that the photoionization of He dominates the heating whereas 
the CNONe IR lines control the cooling. 
We find that an increase of 60 per cent in He abundance or a decrease of 20 per cent in CNONe abundances 
leads to an increase of 200\,K in the equilibrium temperature. It should be mentioned that, because of 
the lack of suitable diagnostic tools, densities of the H-deficient 
inclusions are not well constrained in the current models. This introduces 
some uncertainties in the estimate of the total mass of the cold component.
It is also worth pointing out that in the bi-abundance model, electron temperatures deduced from ratios of IR CELs to optical CELs, such as
the \fneiii~15.6\,$\mu$m/$\lambda$3869 ratio, are expected to be lower than the value derived from the \foiii~optical line ratio, 
consistent with the observations (e.g. Bernard-Salas et al. 2004).

We compare the pressures of the cold inclusions and their surrounding gas in Fig.~16.
On average, the two components have a pressure ratio $P({\rm cold})/P({\rm normal})=0.53\pm0.77$,
which means that the H-deficient knots have a large range of pressure relative to the surrounding gas. 
{\bf This departure from pressure equilibrium could be a consequence of some simplifying assumptions of the model. 
In the model we assume that the H-deficient knots have the same hydrogen number density, 
and the main nebula has a smoothly varying density distribution except the equatorial torus. 
However, the departure is not deleterious to our conclusions.}

The H-deficient knots are thermally stable. If the electron temperature increases, the enhanced thermal 
pressure will force the knots to expand and thus decrease the density, leading to enhanced cooling that will reduce the temperature.
This property may help survival of H-deficient knots during the PN evolution phase.
\begin{figure}
\label{bidensity}  
  \centering
  \epsfig{file=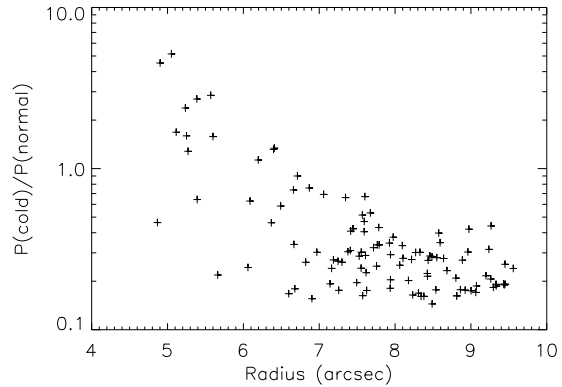, width=8cm,angle=0}
  \caption{ Pressure ratios of cold H-deficient knots and their surrounding gas in the bi-abundance model.}
\end{figure}

In the current bi-abundance model, the cold H-deficient knots are distributed in the nebular inner region (within 10 arcsec)
in order to reproduce the ORL surface brightness distributions yielded by long-slit observations (L2000).
It seems to be a common characteristic of PNe that ORL strengths peak at the nebular centre, e.g.,
PN NGC\,7009 (Luo et al. 2001; Tsamis et al. 2008), PN NGC 6720 (Garnett \& Dinerstein 2001). 
Our modelling also shows that in order to match the strength of the \fneii~ 12.8\,$\mu$m line, which originates mainly from the cold component,
the knots must be located relatively close to the central star in order to avoid over-production of Ne$^+$ ions. 
It is unclear whether this concentration of H-deficient inclusions near the nebular centre is a 
consequence of their lower expansion velocity compared to the main diffuse gas or whether they form from a later evolutionary 
phase of the central star.

Given their high densities, high CNONe abundances and low ionization degrees, the cold knots absorb
hard ionizing photons more efficiently than the diffuse gas, and thus weaken and soften the stellar  radiation fields passing through them.
As a result, the radial shadow tails behind the H-deficient knots have lower temperatures and ionization degrees than their nearby
``unshielded'' gas. These effects are visible in Figs.~12 and 13. The presence of these shadow regions 
is however unlikely to have major effects on ORLs and low-ionization CELs or on $t^2$ values.

Finally, we point out that the current investigation of the properties of H-deficient inclusions can be 
much improved with better IR observations. Restricted by the computer memory, the knots are not spatially resolved at all in the
current  modelling. This rules out the possibility to study structure and properties of optically thick H-deficient knots
as well as effects of knots smaller than the model resolution limit.
In addition, the current work does not take into account the effects of dust grains, a potential heating source 
(e.g. Stasi\'{n}ska \& Szczerba 2001). Given the different physical and chemical conditions, 
the dust grains in H-deficient knots presumably differ in properties
from those in the ``normal'' gas. Future modelling that takes into account the effects of dust 
grains on nebular ionization and thermal structures may be worthwhile.

\begin{table}
  \centering
  \caption{Properties of the H-deficient inclusions in model B.}\label{coldcomponent}

  \begin{tabular}{|l|l|}
    \hline
    $\langle$${\it T}_{\rm e}$$\rangle$ (K)   &815   \\
    $\langle$${\it N}_{\rm H}$$\rangle$ (cm$^{-3}$) &4\,000    \\
    $\langle$${\it N}_{\rm e}$$\rangle$ (cm$^{-3}$)   &6\,680  \\
    {\it M} (M$_{\odot}$)                      &0.0031 \\
    Mass fraction                         &1.3 \% \\
      Number of cells                             &872\\ 
      Cell size                                & $0\farcs25 \times  0\farcs25 \times 0\farcs33$\\
      Filling factor                      & 0.002\\
       He/H                                  &0.50   \\
       C/H                            &0.01177   \\
       N/H                            &0.0150   \\
       O/H                            &0.0440   \\
       Ne/H                           &0.0177   \\
     \hline
 \end{tabular}
\end{table}

\subsection{Size of H-deficient knots}

\begin{figure*}
\centering
\epsfig{file=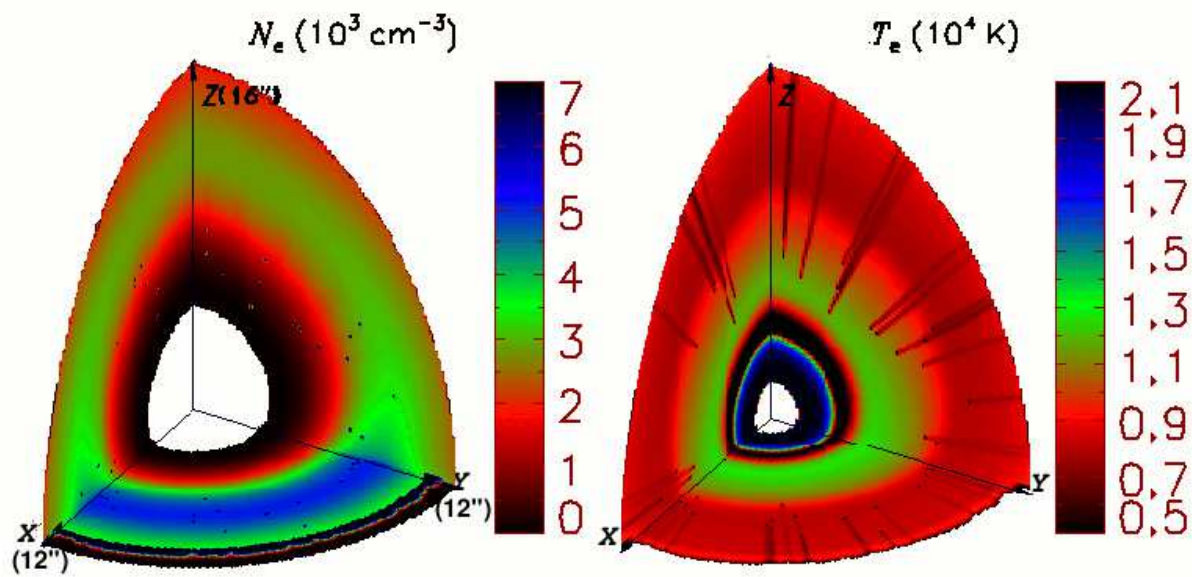, width=16cm}
\caption{Three-dimensional structures of \eld (left panel) and \elt~(right panel) of model B$'$ that
used a higher spatial resolution of 96$^3$ (see text).}
\end{figure*}

\begin{figure*}
\centering
\epsfig{file=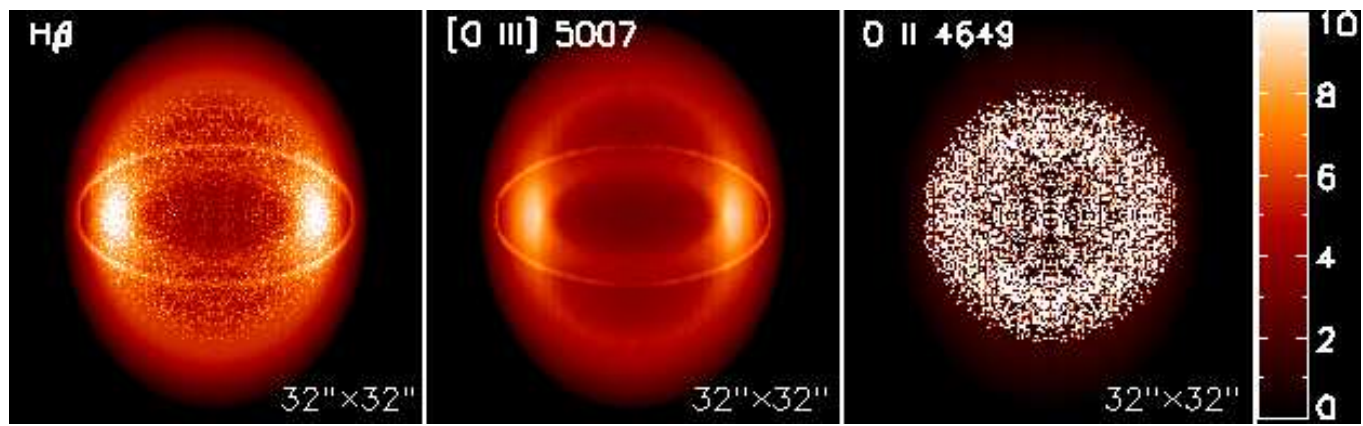 ,width=18cm}
\caption{Projected monochromatic images of \hb~(left panel), \foiii~$\lambda$5007 (central panel) and O~{\sc ii}~$\lambda$4649 (right panel) of
model B$'$ at a viewing angel of $\theta$ = 60 deg.
The units of the color bar are $10^{-13}, 10^{-12}$ and~ $10^{-15}~ {\rm ergs}~ {\rm cm}^{-2}~ {\rm s}^{-1}~ {\rm arcsec}^{-2}$
for the three images, respectively.
}

\end{figure*}

The grainy \hb~image predicted by model B (Fig. ~14) is not consistent with the {\it HST} observation,
suggesting that size of the postulated H-deficient inclusions must be smaller than adopted in this model.
In order to investigate possible effects of assumed knot size on model results,
we have constructed another bi-abundance model of higher resolution using a grid of 96$^3$ cells (model B$'$).
For this new model, all the input parameters are the same as those used in
model B except that the physical size of knots (cells) is reduced by half and the number of knots is increased by a factor of 8.
Fig.~17 shows the electron density and temperature structures predicted by model B$'$. The resultant ORL and CEL images are displayed in Fig.~18.
The predicted total line fluxes as well as the contributions from the
two components are presented in the last three columns of Table~1.
For most lines, the differences between models B and B$'$ are small.
The cold H-deficient knots in model B$'$ have slightly higher ionization degrees than those in model B due to their smaller
physical  size in model B$'$.
There are however significant differences between the two models for the ionic fractions of high-ionization ions 
such as C$^{3+}$, N$^{3+}$, O$^{3+}$ and He$^{2+}$ as well as low ionization ions such as Ne$^+$.
As a result, the predicted fluxes of lines from those ions vary dramatically. 
For example, the \fneii~12.8\,$\mu$m line flux emitted by the cold H-deficient knots in model B$'$
decreases by 50 per cent compared to model B, while the fluxes of \ciii, \niii~ORLs emitted by the cold H-deficient  knots 
increase by 50 per cent, and the fluxes of \heii~lines and the \foiv~25.9\,$\mu$m line are almost doubled.
These results imply that lines such as the \fneii~12.8\,$\mu$m  CEL and \ciii, \niii~ORLs can potentially be used to 
constrain the size of the cold H-deficient inclusions. This is however complicated by the fact that 
intensities of these lines are also affected by the electron density in the cells as well as their spatial distribution.

In an attempt to better constrain the size of the cold knots, the STIS spectra obtained by us have been used 
to deduce the surface brightness distribution of the \cii~$\lambda$4267 line, the strongest ORL detected by STIS. 
The spectra are unfortunately quite noisy and suffer from severe cosmic ray contamination.
Nevertheless, we find the observed surface brightness distribution across the nebula is smoother than 
predicted by model B$'$, suggesting that the postulated cold H-deficient inclusions must have sizes smaller
than $1/6\arcsec$ (i.e. 250 AU at a distance of 1.5 kpc).

\subsection{Elemental abundances}

In Table~5, we list the chemical abundances of NGC\,6153 derived from the bi-abundance model B and those 
in the literature obtained from traditional empirical analyses. For comparison, 
we also give the solar abundances presented by Asplund et al. (2009).  
Table~5 shows that the heavy element abundances deduced from optical/UV CELs, which are systematically lower than 
values yielded by infrared CELs (Pottasch et al. 2003), are in excellent agreement with those of the ``normal'' component; 
whereas the heavy element abundances deduced from infrared CELs are close to the average values of the whole nebula, and are 
much lower than the abundances in the cold component. 
The helium abundance published in the literature, deduced from the empirical method assuming a homogeneous composition,
has been significantly overestimated due to the contamination of cold knots (see also P{\'e}quignot et al. 2002).
If this is also the case for H~{\sc ii} regions, the primordial helium abundance deduced from the standard analyses of helium
recombination lines of metal-poor galaxies will be overestimated. 
The He, C, N, O and Ne abundances in the cold knots are respectively 5, 55, 40, 80 and 100 times higher than those in
the main nebula.

\begin{table}
  \centering
  \caption{Elemental abundances of NGC\,6153.}\begin{tabular}{|l|l|l|l|l|l|l|}
              \hline
              Elem.  & L2000$^a$ & P03$^b$  &B$_{\rm n}$    &B$_{\rm c}$   &B  &Solar$^c$\\
              He      &0.14 & 0.14 & 0.10 &0.50 &0.102  &0.085 \\
              C($-$4) & 2.8 & 6.8  & 3.2  &177  &3.88   &2.692\\
              N($-$4) & 2.3 & 4.8  & 3.8  &150  &4.37   &0.676 \\
              O($-$4) & 5.0 & 8.3  & 5.63 &440  &7.33   &4.898 \\
              Ne($-$4)& 1.7 & 3.1  & 1.76 &177  &2.44   &0.851 \\
              Mg($-$5)&     &      & 3.8  &12.1 &3.83   &3.981  \\
              Si($-$5)&     &      & 3.5  &11.3 &3.53   &3.236 \\
              S($-$5) & 1.6 & 1.9  & 1.75 &5.16 &1.76   &1.318  \\
              Cl($-$7)& 4.2 & 5.6  & 2.35 &10.1 &2.38   &3.162 \\
              Ar($-$6)& 2.7 & 8.5  & 2.9  &11.5 &2.93   &2.512  \\
              \hline
            \end{tabular}
          \begin{description}
       \item[$^a$]  from L2000 for optical/UV CELs.
       \item[$^b$]  from Pottasch et al. (2003).
       \item[$^c$]  from Asplund et al. (2009).
         \end{description}
  \end{table}

\subsection{Possible origins of the postulated  H-deficient inclusions}

Given that adf values measured for PNe are always larger than unity, H-deficient inclusions are 
presumably a genuine feature of PNe. Their presence in PNe is however not predicted by   
current theories of stellar evolution.
While it has been proposed (Iben et al. 1983) that  an evolved star undergoing a very late helium flash 
(the so called ``born-again'' PNe) may harbor H-deficient material, 
the central stars of most PNe exhibiting large adf's are found not to be H-deficient. 
Hydrogen-deficient knots have been clearly detected in the two ``born-again'' PNe, Abell\,30 and Abell\,58. 
On the other hand, Wesson et al. (2003, 2008) found that the H-deficient knots in Abell\,30 and Abell 58 are oxygen-rich, 
in contrast to the expectation of the ``born-again'' scenario.
Their chemical composition has more in common with neon novae than with Sakurai's Object. The latter 
is believed to have recently experienced a final He-shell flash.  
Lau et al. (in preparation) explore the possibility of binary evolution to account for 
the observed high oxygen and neon abundances of the H-deficient knots in Abell\,58, assuming that the knots
are from the ejecta of a neon nova explosion. 
They consider a number of scenarios but none fit perfectly with the observations.
Their best scenario consists of binary stars where the primary ONe white dwarf provided the neon nova explosion  
and the secondary asymptotic giant branch star evolved into the PN. This scenario requires very fine-tuned initial conditions, 
particularly the initial separation, to make the neon nova occur just after the final flash of the secondary asymptotic giant branch 
companion. Therefore, this scenario does not predict a large number of PNe with H-deficient knots. 

Another conjecture is that the H-deficient inclusions are evaporating solid bodies, such as
metal-rich planetesimals formed in planetary disks (Liu et al. 2006). This scenario has been subsequently 
investigated by Henney \& Stasi\'{n}ska  (2010),
who show that the destruction of solid bodies during the short PN phase itself is not able to provide 
enough metal-rich material to explain the observed CEL/ORL abundance discrepancies, due to a low sputtering 
rate.  On the other hand, the sublimation of solid bodies during the final stages of the asymptotic 
giant branch phase, which is much longer than the PN phase, might provide enough metal-rich material and serve as a possible 
mechanism causing the abundance discrepancies. 

Although other explanations to the (moderate) ADF of HIIRs are still possible, 
the H-deficient inclusions may also be present in H~{\sc ii} regions (Tsamis et al. 2003;  Tsamis \& P\'equignot, 2005).
They could be supernovae ejecta in the form of metal-rich droplets that fall back into the interstellar medium 
after a long journey lasting about 10$^8$ yr (Stasi\'{n}ska et al. 2007).
They might also originate from evaporating
proto-planetary disks around newly formed stars, such as the proplyds discovered in the Orion Nebula.

\section{Summary}

We have constructed detailed three-dimensional photoionization models of NGC\,6153 of spherical or ellipsoidal geometry, with and without
H-deficient inclusions. The model results are compared with spectroscopic observations from the UV,  optical to the far IR
as well as narrow-band imaging observations.  
In those models, the main nebula is assumed to be chemically homogeneous and of a ``normal'' composition.
The spherical model can reproduce the strengths of hydrogen and helium recombination lines and of high excitation CELs.
In order to explain the strengths of low-excitation CELs, we have considered an ellipsoidal shell where the  
density decreases from the equator to the poles, supplemented with  an equatorial torus having the same chemical composition
but of a higher density. The chemically homogeneous ellipsoidal model with a torus reproduces the strengths of all CELs 
within the measurement uncertainties, but fails by an order of magnitude for heavy element ORLs.
In all cases,  chemically homogeneous models yield small temperature fluctuations, and are therefore incapable to explain
the large discrepancy between $T_{\rm e}({\rm [O~{\sc III}]})$ and $T_{\rm e}({\rm H~{\sc I}})$.  
In contrast, a bi-abundance model, incorporating small-scale chemical 
inhomogeneities in the form of a small amount of metal-rich inclusions, not only reproduces the strengths of heavy element ORLs and 
the hydrogen Balmer discontinuity, but also improves the overall fitting of CELs.

In our best-fit bi-abundance model, the CNONe abundances in the metal-rich knots are enhanced by 40--100 times compared to 
those of the main nebula. The presence of those metal-rich knots increases the average metal content of the whole 
nebula by 30 per cent. Optical/UV CELs arise mainly
from the main diffuse nebula of ``normal'' composition and thus probe the physical conditions and chemical abundances of this component.
IR CELs originate from both components, and thus yield elemental abundances higher than optical/UV CELs but lower than ORLs.
In addition, temperatures deduced from IR CELs tend to be lower than values deduced from optical/UV CELs.
For \hei~lines, although a large fraction of their fluxes arises from the H-deficient component, given the relatively low abundance 
enhancement factor of about 5, the helium abundance of the normal component remains a fair representation of the average value of the whole nebula.
Nevertheless, in the presence of H-deficient inclusions, the traditional analysis assuming a single uniform chemical composition for
the whole nebula, will greatly overestimate the helium abundances of the normal component. 
One should bear in mind that in the current work we were unable to separate helium emission originating from
the two components to obtain reliable estimates of the helium abundance in each of the two components.
Indeed, we have assumed He/H $=0.1$ for the normal component.
The collisional excitation dominant \hei~$\lambda$~10830 line can, in principle, be used to estimate the helium abundance in 
the normal component. Its usage is however hampered by observational and theoretical uncertainties. 
The electron temperature inside the metal-rich knots is as low as 800\,K. The main coolants are IR fine-structure lines 
from heavy element ions, including the \foiii~52, 88\,$\mu$m, the \fneiii~15.6, 36\,$\mu$m and \fneii~12.8\,$\mu$m lines. 
The metal-rich knots have a high density, and are roughly in pressure equilibrium with the surrounding 
``normal'' gas. The metal-rich knots weaken and soften the UV radiation field. The cometary tails of ``normal'' gas 
behind the knots have relatively low temperatures and ionization degrees. 

We show that the presence of metal-rich knots may lead to significant errors in nebular properties derived 
from the traditional methods assuming a single nebular component. We discuss physical and chemical properties
of the metal-rich knots in NGC\,6153 as well as their possible origins. 
To better constrain the nature of metal-rich knots in PNe, new observations and techniques are 
useful. For example, both imaging and spectroscopic observations of IR fine-structure lines at 
high spatial resolution and sensitivity will certainly shed new insight into the properties of these metal-rich knots.
In future work, we plan to develop more comprehensive
methods to probe these cold knots by involving more observational features (e.g. He~{\sc I} discontinuities; 
Zhang et al. 2005a, 2009) and up-to-date recombination coefficients of ORLs for low-temperature plasma (Fang et al. in preparation).
We will also construct detailed three-dimensional models for more PNe exhibiting large adf's with resolved metal-rich knots, 
such as M~1-42. These are the subjects of further papers.

\vspace{7mm} \noindent {\bf Acknowledgments}
{\bf We would like to thank the referee for the valuable comments, which helped
improve the quality of the paper.}
The modelling was carried out on the SGI Altix330 System at the Department of Astronomy at Peking University, 
HP supercomputer operated by the Center for Computational Science and Engineering at Peking University and 
the Columbia supercomputer operated by the NASA Advanced Supercomputing (NAS) Division at Ames Research Center.
We acknowledge our awards, SMD-08-0633 and SMD-09-1154,
of High-End Computing (HEC) time on NAS and NCCS resources
by the NASA Science Mission Directorate (SMD).
We thank Johnny Chang for his able and generous support
of our research using Columbia.
This research has made use of NASA's Astrophysics Data System Bibliographic Services. 
BE is supported by an STFC Advanced Fellowship.
YZ acknowledges financial support  from the Seed Funding Programme for Basic Research in HKU (200909159007).

\end{document}